\documentclass[preprint,3p]{elsarticle}

\usepackage{graphics}
\usepackage{graphicx}
\usepackage{subfig}
\usepackage{epsfig}

\usepackage{amsmath,amssymb}
\usepackage{amsthm}
\usepackage{float}   
\usepackage{caption}

\usepackage{color}

\usepackage{tikz}
\usetikzlibrary{calc}
\usetikzlibrary{shapes.geometric, arrows}
\usetikzlibrary{arrows.meta}
\tikzset{>={Latex[width=1mm,length=1mm]}}

\tikzstyle{process} = [rectangle, minimum width=2.5cm, minimum height=1cm, text
centered, text width = 3cm, draw=black]
\tikzstyle{decision} = [diamond, minimum width=2.5cm, minimum height=1cm,
aspect=2,inner sep=-0.5ex,text centered, text width = 2.5cm, draw=black]
\tikzstyle{arrow} = [thick,->,>=stealth]
\tikzstyle{line} = [draw, -latex']

\usepackage{rotating}

\usepackage{multirow}
\usepackage[colorlinks]{hyperref}

\biboptions{sort&compress}

\begin{document}

\begin{frontmatter}

\title{Fully-coupled pressure-based algorithm for compressible flows:\\
linearisation and iterative solution strategies}

\author{Fabian Denner\fnref{fn1}}
\ead{fabian.denner@ovgu.de}
\fntext[fn1]{Current address: Chair of Mechanical Process Engineering,
Otto-von-Guericke-Universit\"at Magdeburg, Universit\"atsplatz 2, 39106
Magdeburg, Germany.}

\address{Department of Mechanical
Engineering, Imperial College London, Exhibition Road, London, SW7 2AZ,
United Kingdom}

\begin{abstract}
The impact of different linearisation and iterative solution strategies for
fully-coupled pressure-based algorithms for compressible flows at all speeds is
studied, with the aim of elucidating their impact on the performance of the
numerical algorithm. A fixed-coefficient linearisation and a Newton
linearisation of the transient and advection terms of the governing nonlinear
equations are compared, focusing on test-cases that feature acoustic, shock and
expansion waves. The linearisation and iterative solution strategy applied to
discretise and solve the nonlinear governing equations is found to have a
significant influence on the performance and stability of the numerical
algorithm. The Newton linearisation of the transient terms of the momentum and
energy equations is shown to yield a significantly improved convergence of the
iterative solution algorithm compared to a fixed-coefficient linearisation,
while the linearisation of the advection terms leads to substantial differences
in performance and stability at large Mach numbers and large Courant numbers. It
is further shown that the consistent Newton linearisation of all transient and
advection terms of the governing equations allows, in the context of coupled
pressure-based algorithms, to eliminate all forms of underrelaxation and
provides a clear performance benefit for flows in all Mach number regimes.
\end{abstract}
\end{frontmatter}

\section{Introduction}
The accurate and robust simulation of compressible flows across different Mach
number regimes using the same numerical framework is a widely sought objective
that is notoriously difficult to achieve. The main problems associated with
devising numerical algorithms for flows in all Mach number regimes are finding
suitable discrete formulations that account for the change in mathematical
character of the governing conservation laws, including the related change in
the thermodynamic meaning of pressure and density, and the fully-conservative
discretisation of the governing conservation laws \citep{Anderson1997,
Wesseling2001}.

A straightforward discretisation of the governing conservation laws leads to
{\em density-based} algorithms \citep{Beam1978, MacCormack1982,
Turkel1997}, where density is the solution variable associated with the
conservation of mass, that are particularly suited for flows in which
compressible effects are significant. However, density-based algorithms are
ill-suited for flows with low Mach numbers \citep{Turkel1997, Wesseling2001,
vanderHeul2003, Cordier2012, Miettinen2015}, where the natural coupling between
density and pressure is weak. Consequently, the continuity equation is no longer
effective as a transport equation for density but instead becomes a constraint on the
velocity field. 
The problems associated with density-based algorithms at low Mach numbers and
the desire to be able to simulate flows at all speeds with the same numerical
framework have motivated the development of {\em pressure-based} algorithms
\citep{Harlow1971a, VanDoormaal1987, Chen1991, Acharya2007, Miettinen2015},
in which the continuity equation serves as an equation for pressure, while
density is evaluated explicitly using a suitable equation of state.
In the low Mach number regime, pressure is strongly coupled to
velocity, while the pressure-density coupling is negligible; in the hypersonic
flow regime, pressure is strongly coupled to density, while the
pressure-velocity coupling is negligible. This dual role of pressure facilitates
the success of pressure-based methods in solving flows in all Mach number
regimes \citep{VanDoormaal1987,Acharya2007,Moukalled2016}.
However, pressure-based algorithms exhibit stability and convergence issues when
both the pressure-velocity and the pressure-density coupling are significant
simultaneously, in particular in the transonic flow regime \cite{Wesseling2001},
due to the strong coupling and nonlinearity of the governing equations.

Starting with the seminal work of \citet{Harlow1968, Harlow1971a}, a large
number and varieties of segregated \citep{Issa1986, VanDoormaal1987, Karki1989,
Issa1998, Moukalled2000, Miettinen2015} and coupled \citep{Chen1991,
Demirdzic1993, Karimian1994, Karimian1995, Chen2010, Darwish2014, Xiao2017}
pressure-based algorithms have been proposed for compressible single-phase
flows.
Among the available segregated methods, the class of SIMPLE
\citep{VanDoormaal1987, Karki1989, Issa1998, Moukalled2000} and PISO
\citep{Issa1986, Issa1998, Moukalled2000} methods are most widely used,
providing good performance with low computational resources, in particular
computer memory.
The key shortcoming of segregated methods for compressible flows is the weak
pressure-velocity-density coupling of the discretised governing equations as a
result of the segregated \citep{Kunz1999,Darwish2014}, iterative
predictor-corrector solution procedure, which necessitates underrelaxation of
the discretised equations to reach a converged solution.
The simultaneous solution of the governing equations by coupled methods, in
which all discretised governing equations are solved in a single system of
equations using implicit solution methods, more closely represents their
strongly coupled nature. Although coupled methods typically require larger
computational resources for the solution of the linear system of discretised
governing equations than segregated methods, they benefit from an improved
convergence and robustness \citep{Chen1991,Darwish2014}, in particular on large
computational meshes and for complex flows. Recently, \citet{Xiao2017} proposed
a coupled pressure-based algorithm with a dual-loop solution procedure to
circumvent explicit underrelaxation, featuring an inner iteration loop in which
density is updated assuming the flow is barotropic, with which stable
convergence has been demonstrated for flows in all Mach number regimes
\citep{Xiao2017}.

A point of particular interest when developing numerical algorithms for strongly
nonlinear phenomena, such as compressible flows, is the type of linearisation
applied to the nonlinear governing equations. A well-suited linearisation
strategy can provide a substantial increase in performance and stability of the
numerical algorithm \cite{Dennis1996, Anderson1997, Kunz1999, Wesseling2001}.
Two linearisation methods that are particularly popular and widely applied in
numerical algorithms to predict fluid flows are the {\em fixed-coefficient
linearisation} (or ``lagging" the coefficients) and the {\em Newton
linearisation} (also known as Newton-Raphson method).
In the fixed-coefficient linerisation, only the primary solution variable is
solved implicitly, while all coefficients are computed based on known
information.
For a generic primary variable $\phi$ with its variable coefficient $\alpha$,
the nonlinear term $\alpha^{(n+1)} \,\phi^{(n+1)}$ to be solved, with $n$ the
iteration counter, is approximated as
\begin{equation}
\alpha^{(n+1)} \phi^{(n+1)} \approx \alpha^{(n)} \phi^{(n+1)}  \ ,
\label{eq:fixedCoeffGeneric}
\end{equation}
where superscript $(n)$ denotes the most recent available solution. In the
context of pressure-based algorithms, arguments for a fixed-coefficient
linearisation are its easy implementation, and that it is not necessary to
treat the fluxes and the density implicitly as a function of one of the primary
solution variables.
The Newton linearisation is an often chosen alternative to the fixed-coefficient
linearisation, see {\em e.g.}~\citep{Darbandi2008, Darwish2014, Xiao2017}, providing superior
convergence rates and stability of the solution algorithm
\citep{Wesseling2001}, as for instance demonstrated by \citet{Kunz1999} in the
context of a coupled pressure-based multi-fluid Euler-Euler method.
Applying a Newton linearisation, the nonlinear term $\alpha^{(n+1)}
\,\phi^{(n+1)}$ is approximated as
\begin{equation}
\begin{split}
\alpha^{(n+1)} \phi^{(n+1)} & \approx \alpha^{(n)} \phi^{(n)} +
\left( \alpha^{(n+1)} - \alpha^{(n)} \right) \left. \frac{\partial \alpha
\phi}{\partial \alpha} \right|^{(n)} + \left(\phi^{(n+1)} - \phi^{(n)}
\right) \left. \frac{\partial \alpha \phi}{\partial \phi} \right|^{(n)} \\
 & = \alpha^{(n)} \phi^{(n+1)} + \alpha^{(n+1)} \phi^{(n)} - \alpha^{(n)}
 \,\phi^{(n)} \ .
\end{split}
\label{eq:NewtonGeneric}
\end{equation}
Arguments in support of the Newton linearisation for simulating compressible
flows typically point to a suitable treatment and smooth transition from
elliptic/parabolic to hyperbolic behaviour of the governing equations
\citep{Karimian1994, Darbandi2004, Xiao2017}, in particular the continuity
equation, in different Mach number regimes, as well as an implicit contribution
of additional active flow-dependent variables, such as the fluxes. Despite the
often stated importance of the applied linearisation for the performance and
robustness of the solution algorithm in the relevant literature, notably
textbooks \citep{Anderson1997,Wesseling2001, Ferziger2002}, a systematic study
of the linearisation for pressure-based algorithms for compressible flows has
not been published to date.

In this article, the linearisation of the governing equations as well as the
iterative solution strategy for a fully-coupled pressure-based algorithm for the
simulation of flows at all speeds, based on the framework proposed by
\citet{Xiao2017}, is studied. Considering the fixed-coefficient and Newton
linearisations, the different possible linearisation strategies for each term of
the governing equations are studied and the resulting performance and stability
of the numerical algorithm are compared using representative test-cases in all
Mach number regimes, including the propagation of acoustic waves, shock tubes
and supersonic flows over a forward-facing step and a circular cone. The
presented results demonstrate subtle differences between the considered
linearisation strategies and highlight the importance of a careful linearisation
of the governing nonlinear equations. The execution times for all presented
simulations are given together with the used computational hardware, as a
reference for future algorithm development and comparisons.

The governing equations are briefly introduced in Section \ref{sec:governingEq}
and the applied numerical framework is presented in Section \ref{sec:numerics}.
The considered linearisation techniques and solution procedures are discussed in
Section \ref{sec:linearisation} and the results of representative test-cases are
presented in Section \ref{sec:results}. The findings are summarised and the
article is concluded in Section \ref{sec:conclusions}.

\section{Governing equations}
\label{sec:governingEq}
The considered compressible flows of an inviscid fluid are
governed by the continuity, momentum and energy equations, given as (using the
Einstein notation)
\begin{align}
\frac{\partial \rho}{\partial t} + \frac{\partial \rho u_i}{\partial x_i} & = 
0 \ , \label{eq:continuity} \\
\frac{\partial \rho u_j}{\partial t} + \frac{\partial \rho u_i u_j}{\partial
x_i} & = - \frac{\partial p}{\partial x_j}  
\ ,
\label{eq:euler} \\
\frac{\partial \rho h}{\partial t} + \frac{\partial \rho u_i h}{\partial x_i} &
= 
\frac{\partial p}{\partial t}  
\ , \label{eq:energy}
\end{align}
respectively, where $t$ is time, $\boldsymbol{x}$ the Cartesian coordinates,
$\rho$ is the density, $\boldsymbol{u}$ is the velocity vector, $p$ is the pressure and $h =
c_p \, T + \boldsymbol{u}^2/2$ is the specific total enthalpy, with $c_p$ the
specific isobaric heat capacity and $T$ the temperature. For simplicity, but
without loss of generality, viscous stresses, heat conduction and external
forces are neglected in this study. The system of governing equations is closed
by the ideal gas equation of state
\begin{equation}
\rho = \frac{p }{(\gamma-1) \, c_v \, T} \ , \label{eq:eos}
\end{equation}
where $c_v$ is the specific isochoric heat capacity and $\gamma = c_p/c_v$ is
the heat capacity ratio. The speed of sound is given as
\begin{equation}
a = \sqrt{\frac{\gamma \, p}{\rho}} \ . \label{eq:soundSpeed}
\end{equation}

\section{Numerical framework}
\label{sec:numerics}
A coupled pressure-based finite-volume framework for compressible flows, based
on the numerical framework of \citet{Xiao2017}, is employed to
solve the governing equations. This numerical framework is founded on a
collocated variable arrangement, is applicable to unstructured meshes and does
not apply any explicit underrelaxation to the iterative solution
algorithm. In this section, the discretisation and implementation of the
governing equations is explained, focusing on the methods and ingredients
relevant to this study. Further details on the applied finite-volume framework
can be found in previous work \citep{Xiao2017}.
The considered linearisation and solution strategies are discussed in Section
\ref{sec:linearisation}. 

\subsection{Spatial and temporal discretisation}
\begin{figure}
\begin{center}
\begin{tikzpicture}[scale=1]
\draw [->] (2,1) -- (2.6,1);
\node [below] at (2.6,1) {$\boldsymbol{n}_f$};
\draw [thick] (0,0) -- (4,0);
\draw [thick] (4,0) -- (4,2);
\draw [thick] (4,2) -- (0,2);
\draw [thick] (0,0) -- (0,2);
\draw [thick] (2,0) -- (2,2);
\draw [fill] (1,1) circle [radius=0.05];
\node [above] at (1,1) {$P$};
\draw [fill] (3,1) circle [radius=0.05];
\node [above] at (3,1) {$Q$};
\draw [fill] (2,1) circle [radius=0.05];
\node at (2.2,1.2) {$f$};
\end{tikzpicture} \label{fig:discGeoGeneral}
\caption{Schematic illustration of cell $P$ with its neighbour cell $Q$ and
the shared face $f$, where $\boldsymbol{n}_f$ is the unit normal
vector of face $f$ (outward pointing with respect to cell $P$).}
\label{fig:discGeo}
\end{center}
\end{figure}
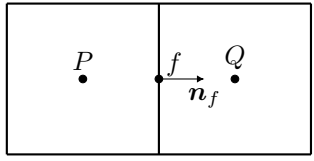

The central differencing scheme is applied for the interpolation from cell
centres to face centres of variables that are not advected, given for a general
flow variable $\phi$ at face $f$, see Fig.~\ref{fig:discGeo}, as
\begin{equation}
\overline{\phi}_f = \frac{\phi_P + \phi_Q}{2} \ .
\label{eq:cellFaceInterp}
\end{equation} 
Advected variables are interpolated to face centres using the Minmod scheme
\citep{Roe1986}, with the face value following as 
\begin{equation}
\tilde{\phi}_f = \phi_U + \frac{\xi_f}{2} (\phi_D-\phi_U) \ ,
\end{equation} 
where subscripts $U$ and $D$ denote the upwind and downwind cells, and $\xi_f$
is the flux limiter. Other TVD schemes
would be equally applicable but are not considered as part of this study.

The First-Order Backward Euler and the Second-Order Backward
Euler schemes are applied for the discretisation of transient terms, given for
cell $P$ as
\begin{equation}
\int_{V_P} \frac{\partial \phi}{\partial t} \ dV \approx \frac{\phi_P -
\phi_P^{(t-\Delta t)}}{\Delta t} \, V_P  \label{eq:foe}
\end{equation}
and 
\begin{equation}
\int_{V_P} \frac{\partial \phi}{\partial t} \ dV \approx \frac{3 \phi_P - 4
\phi_P^{(t-\Delta t)} + \phi_P^{(t-2\Delta t)}}{2 \Delta t} \, V_P \ ,
\end{equation}
respectively, where  $\Delta t$ is the applied time-step, superscripts
$(t-\Delta t)$  and $(t-2\Delta t)$ denote values of the previous time-level and
the previous-previous time-level, respectively, and $V_P$ is the volume of cell
$P$. For simplicity, the discretised governing equations are presented below
using the First-Order Backward Euler scheme, but the Second-Order
Backward Euler scheme is also applied as part of this study. In the
interest of consistency, all transient terms of the governing equations are
always discretised with identical schemes.

\subsection{Advecting velocity}
At cell faces $f$, an advecting velocity $\vartheta_f=\boldsymbol{u}_f
\boldsymbol{n}_f$ is defined using the momentum-weighted interpolation method,
with $\boldsymbol{n}_f$ the unit normal vector of face $f$. This advecting
velocity takes the role of flux-velocity in the discretised advection terms of
the governing equation. Following the work of \citet{Xiao2017}, the advecting
velocity $\vartheta_f$ at face $f$ is defined as
\begin{equation}
\vartheta_f \approx \overline{u}_{f,i} \, n_{f,i}
- \hat{d}_f \left[ \left. \frac{\partial p}{\partial x_i} \right|_f  n_{f,i} -
\frac{1}{2} \left( \left.
\frac{\partial p}{\partial x_i} \right|_P +
\left. \frac{\partial p}{\partial x_i} \right|_Q
\right) n_{f,i} \right]  
+ \hat{d}_f \frac{\rho_f^{(t-\Delta t)}}{\Delta
t} \left(\vartheta^{(t-\Delta t)}_{f} - \overline{u}^{(t-\Delta t)}_{f,i} \,
n_{f,i} \right) \, ,
\label{eq:MWIfull}
\end{equation}
where $\overline{\boldsymbol{u}}_f$ is obtained by linear interpolation from the
values at the adjacent cell centres and
\begin{equation}
\left. \frac{\partial p}{\partial x_i} \right|_f n_{f,i} \approx \frac{p_Q -
p_P}{\Delta x} \ .
\end{equation}
The coefficient $\hat{d}_f$ follows directly from the
coefficients of the advection terms (and, if considered, viscous stress terms)
of the momentum equations, see for instance \citep{Denner2018b}. 
This formulation of the advecting velocity provides a robust pressure-velocity
coupling at all Mach numbers \cite{Xiao2017}. For low Mach numbers and incompressible flows, the
pressure term acts as a low-pass filter on high-order derivatives of pressure
\citep{Demirdzic1995,Wesseling2001,Ferziger2002}, because
\begin{equation}
\left. \frac{\partial p}{\partial x_i} \right|_f - \frac{1}{2}
\left( \left. \frac{\partial p}{\partial x_i} \right|_P +
\left. \frac{\partial p}{\partial x_i} \right|_Q
\right) \propto \left. \frac{\partial^3 p}{\partial x_i^3}
\right|_f \ ,
\end{equation}
which damps pressure oscillations arising as a result of pressure-velocity
decoupling in a collocated variable arrangement.

\subsection{Discretised governing equations}
Applying the First-Order Backward Euler scheme (chosen here for demonstration),
given by Eq.~(\ref{eq:foe}), for the discretisation of the transient terms and
the advecting velocity $\vartheta_f$, given by Eq.~(\ref{eq:MWIfull}), in the
advection terms, the discretised continuity equation (\ref{eq:continuity})
for mesh cell $P$ is given as
\begin{equation}
\frac{\rho_P^{(n+1)} -\rho_P^{(t-\Delta t)}}{\Delta
t} \, V_P + \sum_f \tilde{\rho}_f \vartheta_{f} A_f =
0 \ , \label{eq:discConti} 
\end{equation}
where $A_f$ is the area of face $f$. For all results presented as part of
this study, the cell-centred density $\rho_P^{(n+1)}$ in the transient term of
the continuity equation is formulated as an implicit function of pressure $p$, given
as
\begin{equation}
\rho^{(n+1)}_P = \frac{p^{(n+1)}_P}{(\gamma - 1) \, c_v \,
T_P}  \ , \label{eq:discEOS}
\end{equation}
where $T$ is the most recent available temperature value, which is
dependent on the applied solution procedure and is detailed in Section
\ref{sec:solutionStrategy}. The linearisation of the advection term of
Eq.~(\ref{eq:discConti}) is discussed in Section \ref{sec:linearisationConti}.
The discretised momentum equations (\ref{eq:euler}) and energy equation
(\ref{eq:energy}) for mesh cell $P$ follow in a similar manner as
\begin{align}
\frac{\rho_P u_{P,j} - \rho_P^{(t-\Delta t)} u_{P,j}^{(t-\Delta t)}}{ \Delta t}
\, V_P + \sum_f \tilde{\rho}_f \vartheta_f \tilde{u}_{f,j} A_f & = - \sum_f \overline{p}_f
n_{f,j} A_f \ , \label{eq:discEuler} \\
\frac{\rho_P h_P - \rho_P^{(t-\Delta t)} h_P^{(t-\Delta t)}}{\Delta t} \, V_P +
\sum_f \tilde{\rho}_f \vartheta_f \tilde{h}_f A_f & = \frac{p_P - p_P^{(t-\Delta
t)}}{\Delta t} \, V_P \ ,
\label{eq:discEnergy}
\end{align}
respectively. The linearisation of the transient and advection terms of
Eqs.~(\ref{eq:discEuler}) and (\ref{eq:discEnergy}) is discussed in Section
\ref{sec:linMomEnergy}.

Note that the continuity, momentum and energy equations are formulated
conservative in $\rho$, $\rho \boldsymbol{u}$ and $\rho h$, respectively, but
are solved for the primary variables $p$, $\boldsymbol{u}$ and $h$, with $\rho$
given by Eq.~(\ref{eq:eos}). All cell-centred values of $p$, $\boldsymbol{u}$
and $h$ arising in the discretised governing equations are treated implicitly,
which is further discussed in Section \ref{sec:linearisation}, and the same
advecting velocity $\vartheta_f$ is applied in the discretised governing
equations to ensure a consistent formulation of the fluxes.
\citet{VanDoormaal1987} and subsequent studies \citep{Chen1991, Karimian1994,
Darwish2014, Xiao2017, Denner2018b} demonstrated that choosing primitive
variables instead of conserved variables as primary solution variables does not affect the
conservative properties of the governing equations, if a consistent
discretisation is applied. Although the continuity equation acts as a constraint
on the pressure field, the resulting density and velocity fields, through the
coupling with the momentum equations and the applied equation of state, satisfy
the conservation of mass in all Mach number regimes \citep{VanDoormaal1987}. The
converged system of nonlinear governing equations, thus, satisfies the governing
conservation laws on the discrete level.

\subsection{Linear system of equations}
The discretised governing equations are solved in a single linear system of
equations, $\boldsymbol{A} \boldsymbol{\phi} = \boldsymbol{b}$, which for a
three-dimensional flow is given as
\begin{equation}
\begin{pmatrix}
{\boldsymbol{A}}^u_{\rho u} & 
{\boldsymbol{A}}^v_{\rho u} &
{\boldsymbol{A}}^w_{\rho u} & 
{\boldsymbol{A}}^p_{\rho u} &
{\boldsymbol{0}} \\
{\boldsymbol{A}}^u_{\rho v} & 
{\boldsymbol{A}}^v_{\rho v} &
{\boldsymbol{A}}^w_{\rho v} & 
{\boldsymbol{A}}^p_{\rho v} &
{\boldsymbol{0}} \\
{\boldsymbol{A}}^u_{\rho w} & 
{\boldsymbol{A}}^v_{\rho w} &
{\boldsymbol{A}}^w_{\rho w} & 
{\boldsymbol{A}}^p_{\rho w} &
{\boldsymbol{0}} \\
{\boldsymbol{A}}^u_{\rho \phantom{h}} & 
{\boldsymbol{A}}^v_{\rho \phantom{h}} &
{\boldsymbol{A}}^w_{\rho \phantom{h}} & 
{\boldsymbol{A}}^p_{\rho \phantom{h}} &
{\boldsymbol{0}} \\
{\boldsymbol{A}}^u_{\rho h} & 
{\boldsymbol{A}}^v_{\rho h} &
{\boldsymbol{A}}^w_{\rho h} & 
{\boldsymbol{A}}^p_{\rho h} &
{\boldsymbol{A}}^h_{\rho h}
\end{pmatrix} \cdot
\begin{pmatrix}
\boldsymbol{\phi}^u \\
\boldsymbol{\phi}^v \\
\boldsymbol{\phi}^w \\
\boldsymbol{\phi}^p \\
\boldsymbol{\phi}^h
\end{pmatrix} =
\begin{pmatrix}
\boldsymbol{b}_{\rho u} \\
\boldsymbol{b}_{\rho v} \\
\boldsymbol{b}_{\rho w} \\
\boldsymbol{b}_{\rho \phantom{h}} \\
\boldsymbol{b}_{\rho h}
\end{pmatrix}
 \ ,
\label{eq:eqsys}
\end{equation}
where $\boldsymbol{A}^\chi_\zeta$, with $\zeta$ the conserved quantity of a
given governing equation and $\chi$ the solution variable, are the coefficient
submatrices of the momentum equations ($\zeta = \{\rho u, \rho v, \rho w\}$),
the continuity equation ($\zeta = \rho$) and the energy equation ($\zeta = \rho
h$).
The vectors $\boldsymbol{\phi}^\chi$ and $\boldsymbol{b}_\zeta$ are the solution
subvectors and right-hand side subvectors, respectively.
Note that, contrary to the work of \citet{Xiao2017}, the discretised energy
equation is solved together with the momentum and continuity equations in the
linear system of equations (\ref{eq:eqsys}), to facilitate the implicit coupling
provided by some of the studied linearisation strategies.
The applied linearisation strategies, presented in Sections
\ref{sec:linearisationConti} and \ref{sec:linMomEnergy}, determine the
sparseness of the equation system. For the results presented in this study, the
system of governing equations (\ref{eq:eqsys}) is preconditioned and solved
using the {\em Block Jacobi} preconditioner and {\em BiCGStab} solver of the
PETSc library \citep{Balay1997, petsc-web-page, petsc-user-ref}, respectively.
The equation system (\ref{eq:eqsys}) has converged if \citep{petsc-user-ref}
\begin{equation}
\| \boldsymbol{A}^{(n)} \boldsymbol{\phi}^{(n+1)} -
\boldsymbol{b}^{(n)} \| < \eta \, \| \boldsymbol{b}^{(n)} \|
\ , \label{eq:linConv}
\end{equation}
where $\eta$ is the predefined solution tolerance and $\| \cdot \|$ denotes the
$L_2$-norm.

\section{Linearisation and iterative solution strategies}
\label{sec:linearisation}
Different linearisation strategies are devised by applying the fixed-coefficient
linearisation, Eq.~(\ref{eq:fixedCoeffGeneric}), and the Newton linearisation,
Eq.~(\ref{eq:NewtonGeneric}), in different combinations to the various nonlinear
transient terms and advection terms of the governing equations.
The considered linearisation strategies are presented in Sections
\ref{sec:linearisationConti} and \ref{sec:linMomEnergy}, and the applied
iterative single-loop and dual-loop solution procedures are discussed in Section
\ref{sec:solutionStrategy}.

\subsection{Linearisation of the continuity equation}
\label{sec:linearisationConti}
The linearisation of the advection term of the continuity equation
(\ref{eq:discConti}) has been discussed in several previous studies, see {\em
e.g.}~\citep{VanDoormaal1987, Issa1998, Karimian1994, Xiao2017}. The general
consensus is that a Newton linearisation for this term is preferable over the
fixed-coefficient linearisation, as it provides a smooth transition from the
elliptic equation for pressure in the incompressible limit ($M\rightarrow 0$) to
the hyperbolic nature of the continuity equation for supersonic flows ($M > 1$).
With different linearisations applied to the term $\rho_f \vartheta_f$, the
discretised continuity equation (\ref{eq:discConti}) becomes
\begin{equation}
\frac{\rho_P^{(n+1)} -\rho_P^{(t-\Delta t)}}{\Delta t} \, V_P + \sum_f
\tilde{\rho}^{(n)}_f \vartheta_{f}^{(n+1)} A_f = 0
\end{equation}
with the fixed-coefficient linearisation, and
\begin{equation}
\frac{\rho_P^{(n+1)} -\rho_P^{(t-\Delta t)}}{\Delta t} \, V_P + \sum_f
\left( \tilde{\rho}^{(n)}_f \vartheta_{f}^{(n+1)} + \tilde{\rho}^{(n+1)}_f
\vartheta_{f}^{(n)} - \tilde{\rho}^{(n)}_f \vartheta_{f}^{(n)} \right) A_f = 0
\end{equation}
with the Newton linearisation, where $\rho_P^{(n+1)}$ is given by
Eq.~(\ref{eq:discEOS}) and $\tilde{\rho}^{(n+1)}_f$ is given as
\begin{equation}
\tilde{\rho}_f^{(n+1)} = \rho_U^{(n+1)} + \frac{\xi_f}{2} \left(\rho_D^{(n+1)} -
\rho_U^{(n+1)} \right) \ , \label{eq:discEOSFace}
\end{equation}
with $\rho_U^{(n+1)}$ and $\rho_D^{(n+1)}$ evaluated by Eq.~(\ref{eq:discEOS}).
The implicit advecting velocity $\vartheta_f^{(n+1)}$ is given as
\begin{equation}
\begin{split}
\vartheta_f^{(n+1)} \approx \overline{u}_{f,i}^{(n+1)} \, n_{f,i}
& - \hat{d}_f \left[ \frac{p_Q^{(n+1)} - p_P^{(n+1)}}{\Delta x} -
\frac{1}{2} \left( \left. \frac{\partial p}{\partial x_i} \right|_P^{(n)}  + \left.
\frac{\partial p}{\partial x_i} \right|_Q^{(n)} \right) n_{f,i} \right] \\ & +
\hat{d}_f \frac{\rho^{(t-\Delta t)}_f}{\Delta t}
\left(\vartheta^{(t-\Delta t)}_{f} - \overline{u}^{(t-\Delta t)}_{f,i} \,
n_{f,i} \right) \ ,
\label{eq:MWIfullDisc}
\end{split}
\end{equation}
where the cell-centred values of velocity and pressure are solved for
implicitly, while the cell-centred pressure gradients are deferred. Preliminary
studies have shown no significant differences in performance or stability for
the considered test-cases when the cell-centred pressure gradients in
Eq.~(\ref{eq:MWIfullDisc}) were instead treated implicitly; hence an implicit
treatment of the cell-centred pressure gradients in Eq.~(\ref{eq:MWIfullDisc})
is not considered as part of this study.

For the Newton linearisation, $\tilde{\rho}_f^{(n)} \vartheta_f^{(n+1)}$
dominates over $\tilde{\rho}_f^{(n+1)} \vartheta_f^{(n)}$ for low Mach numbers,
whereas $\tilde{\rho}_f^{(n+1)} \vartheta_f^{(n)}$ dominates over
$\tilde{\rho}_f^{(n)} \vartheta_f^{(n+1)}$ in the hypersonic regime. Therefore,
the fixed-coefficient linearisation, which is derived from a pressure-based
numerical framework for incompressible flows, as discussed in \citep{Xiao2017},
is expected to yield a very limited performance and stability for flows with large
Mach numbers. Note that an implicit treatment of the advecting velocity is
essential for a robust pressure-velocity coupling at low Mach numbers
\cite{Xiao2017} and, hence, the fixed-coefficient linearisation with density as
the implicit variable and the advecting velocity as the deferred coefficient is
not considered in this study.

\subsection{Linearisation of the momentum and energy equations}
\label{sec:linMomEnergy}
The momentum and energy equations offer more options for linearisation than the
continuity equation because of the additional primary variable $\phi$, {\em
i.e.}~velocity $\boldsymbol{u}$ in the momentum equations (\ref{eq:discEuler})
and specific total enthalpy $h$ in the energy equation (\ref{eq:discEnergy}).
The transient term $\partial \rho u_j / \partial t$ of the momentum equations
(\ref{eq:discEuler}) and the transient term $\partial \rho h / \partial t$ of
the energy equation (\ref{eq:discEnergy}) are linearised with the
fixed-coefficient linearisation
\begin{equation}
\rho_P \phi_P = \rho_P^{(n)} \phi_P^{(n+1)} \ , 
\end{equation}
or the Newton linearisation
\begin{equation}
\rho_P \phi_P = \rho_P^{(n)} \phi_P^{(n+1)} + \rho_P^{(n+1)} \phi_P^{(n)} -
\rho_P^{(n)} \phi_P^{(n)} \ ,
\end{equation}
with $\rho_P^{(n+1)}$ given by Eq.~(\ref{eq:discEOS}). Because pressure is a
primary solution variable in all governing equations, no additional non-zero
matrix coefficients arise when the density is treated implicitly as a function
of pressure.

Applying different combinations of the fixed-coefficient and Newton
linearisations, four different linearisation strategies can be devised
for the advection terms of the momentum equations
(\ref{eq:discEuler}) and energy equation (\ref{eq:discEnergy}):
\begin{itemize}
  \item Fixed-coefficient linearisation, where only the primary variable is
  treated implicitly,
  \begin{equation}
  \tilde{\rho}_f \vartheta_f \tilde{\phi}_f =
  \tilde{\rho}_f^{(n)} \vartheta_f^{(n)} \tilde{\phi}_f^{(n+1)} \ ,
  \end{equation}
  \item Newton linearisation with respect to the density ($\rho$-Newton
  linearisation),
  \begin{equation}
  \tilde{\rho}_f \vartheta_f \tilde{\phi}_f =
  \tilde{\rho}_f^{(n)} \vartheta_f^{(n)} \tilde{\phi}_f^{(n+1)}
  + \tilde{\rho}_f^{(n+1)} \vartheta_f^{(n)} \tilde{\phi}_f^{(n)} -
  \tilde{\rho}_f^{(n)} \vartheta_f^{(n)} \tilde{\phi}_f^{(n)} \ ,
  \end{equation}
  \item Newton linearisation with respect to the advecting velocity
  ($\vartheta$-Newton linearisation),
  \begin{equation}
  \tilde{\rho}_f \vartheta_f \tilde{\phi}_f = \tilde{\rho}_f^{(n)}
  \vartheta_f^{(n)} \tilde{\phi}_f^{(n+1)} + \tilde{\rho}_f^{(n)}
  \vartheta_f^{(n+1)} \tilde{\phi}_f^{(n)} - \tilde{\rho}_f^{(n)}
  \vartheta_f^{(n)} \tilde{\phi}_f^{(n)}\ ,
  \end{equation}
  \item Full-Newton linearisation by combining the $\rho$-Newton and
  $\vartheta$-Newton linearisations,
  \begin{equation}
  \tilde{\rho}_f \vartheta_f \tilde{\phi}_f  =
  \tilde{\rho}_f^{(n)} \vartheta_f^{(n)} \tilde{\phi}_f^{(n+1)} +
  \tilde{\rho}_f^{(n)} \vartheta_f^{(n+1)} \tilde{\phi}_f^{(n)} + 
  \tilde{\rho}_f^{(n+1)} \vartheta_f^{(n)} \tilde{\phi}_f^{(n)} - 2
  \tilde{\rho}_f^{(n)} \vartheta_f^{(n)} \tilde{\phi}_f^{(n)} \ .
\end{equation}
\end{itemize}
The fully linearised momentum equations (\ref{eq:discEuler}) and energy equation
(\ref{eq:discEnergy}) follow as
\begin{equation}
\begin{split}
& \left(\overbrace{\underbrace{\rho_P^{(n)}
u_{P,j}^{(n+1)}}_\textup{fixed-coeff.} + \rho_P^{(n+1)} u_{P,j}^{(n)} -
\rho_P^{(n)} u_{P,j}^{(n)}}^\textup{Newton} - \rho_P^{(t-\Delta t)}
u_{P,j}^{(t-\Delta t)} \right) \frac{V_P}{ \Delta t} + \sum_f
\overline{p}_f^{(n+1)} n_{f,j} A_f \\
& + \sum_f \left(\underbrace{
\rlap{$\underbrace{\phantom{\tilde{\rho}_f^{(n+1)} \vartheta_f^{(n)}
\tilde{u}_{f,j}^{(n)} - \tilde{\rho}_f^{(n)} \vartheta_f^{(n)}
\tilde{u}_{f,j}^{(n)} +
\tilde{\rho}_f^{(n)} \vartheta_f^{(n)}
\tilde{u}_{f,j}^{(n+1)}}}_\textup{$\rho$-Newton}$} \tilde{\rho}_f^{(n+1)}
\vartheta_f^{(n)} \tilde{u}_{f,j}^{(n)} - \tilde{\rho}_f^{(n)} \vartheta_f^{(n)}
\tilde{u}_{f,j}^{(n)} + \overbrace{\overbrace{\tilde{\rho}_f^{(n)}
\vartheta_f^{(n)} \tilde{u}_{f,j}^{(n+1)}}^\textup{fixed-coeff.} +
\tilde{\rho}_f^{(n)} \vartheta_f^{(n+1)} \tilde{u}_{f,j}^{(n)} -
\tilde{\rho}_f^{(n)} \vartheta_f^{(n)}
\tilde{u}_{f,j}^{(n)}}^\textup{$\vartheta$-Newton}}_\textup{full-Newton} \right)
A_f = 0 \label{eq:discEulerLin}
\end{split}
\end{equation}
and
\begin{equation}
\begin{split}
& \left(\overbrace{\underbrace{\rho_P^{(n)}
h_P^{(n+1)}}_\textup{fixed-coeff.} + \rho_P^{(n+1)} h_P^{(n)} -
\rho_P^{(n)} h_P^{(n)}}^\textup{Newton} - \rho_P^{(t-\Delta t)}
h_P^{(t-\Delta t)} \right) \frac{V_P}{ \Delta t} - \left(p_P^{(n+1)} -
p_P^{(t-\Delta t)}\right) \frac{V_P}{\Delta t}  \\
& + \sum_f \left(\underbrace{
\rlap{$\underbrace{\phantom{\tilde{\rho}_f^{(n+1)} \vartheta_f^{(n)}
\tilde{h}_{f}^{(n)} - \tilde{\rho}_f^{(n)} \vartheta_f^{(n)}
\tilde{h}_{f}^{(n)} +
\tilde{\rho}_f^{(n)} \vartheta_f^{(n)}
\tilde{h}_{f}^{(n+1)}}}_\textup{$\rho$-Newton}$} \tilde{\rho}_f^{(n+1)}
\vartheta_f^{(n)} \tilde{h}_{f}^{(n)} - \tilde{\rho}_f^{(n)} \vartheta_f^{(n)}
\tilde{h}_{f}^{(n)} + \overbrace{\overbrace{\tilde{\rho}_f^{(n)}
\vartheta_f^{(n)} \tilde{h}_{f}^{(n+1)}}^\textup{fixed-coeff.} +
\tilde{\rho}_f^{(n)} \vartheta_f^{(n+1)} \tilde{h}_{f}^{(n)} -
\tilde{\rho}_f^{(n)} \vartheta_f^{(n)}
\tilde{h}_{f}^{(n)}}^\textup{$\vartheta$-Newton}}_\textup{full-Newton} \right)
A_f = 0  \ , \label{eq:discEnergyLin}
\end{split}
\end{equation}
respectively, where the braces indicate which terms are part of the various 
linearisation strategies. Note that, for convenience of presentation, the
pressure terms appearing on the right hand-side of Eqs.~(\ref{eq:discEuler}) and
(\ref{eq:discEnergy}) have been moved to the left-hand side of
Eqs.~(\ref{eq:discEulerLin}) and (\ref{eq:discEnergyLin}).

The $\rho$-Newton linearisation has previously been applied to the momentum
equations by \citet{VanDoormaal1987} and the $\vartheta$-Newton linearisation
has been considered by \citet{Darbandi2004}, who reported an improved
performance of their shock-capturing method. The fixed-coefficient linearisation
and the $\rho$-Newton linearisation result in no additional non-zero matrix
coefficients, since pressure is treated implicitly in all governing equations,
while the $\vartheta$-Newton linearisation yields additional non-zero entries in
the coefficient matrix for all velocity components. Hence, an appreciable
acceleration of convergence has to be achieved with the $\vartheta$- and
full-Newton linearisations to gain an overall performance benefit.

\subsection{Iterative solution procedure}
\label{sec:solutionStrategy}
An {\em inexact Newton} method \cite{Dembo1982} is applied to solve the
nonlinear governing equations, performing nonlinear iterations in which the
deferred variables are updated based on the latest result obtained from solving
equation system (\ref{eq:eqsys}). This iterative procedure continues until,
after updating $\boldsymbol{A}^{(n+1)} \leftarrow \boldsymbol{A} (
\boldsymbol{\phi}^{(n+1)} )$ and $\boldsymbol{b}^{(n+1)} \leftarrow
\boldsymbol{b} ( \boldsymbol{\phi}^{(n+1)} )$, the $L_2$-norm of the residual
vector $\boldsymbol{r}$ of the equation system satisfies
\begin{equation}
\| \boldsymbol{r}^{(n+1)} \| = \frac{\| \boldsymbol{A}^{(n+1)}
 \boldsymbol{\phi}^{(n+1)} -
\boldsymbol{b}^{(n+1)} \|}{ \| \boldsymbol{b}^{(n+1)} \| \, \Theta } <
\eta  \ ,
\label{eq:nonlinConv}
\end{equation}
where $\Theta = \sqrt{N_r}$ is a scaling factor and $N_r$ is the size of
$\boldsymbol{r}$. 

\begin{figure}[t]
\begin{center}
\begin{small}
\subfloat[Single-loop solution procedure]
{\begin{tikzpicture}[node distance=1.5cm]
\node (pro0) [process] {Update previous time-levels};
\node (pro1) [process, below of=pro0] {Assemble and solve 
Eq.~(\ref{eq:eqsys})}; 
\node (pro2) [process, below of=pro1] {Update $T^{(n+1)}$ from $h^{(n+1)}$};
\node (pro3) [process, below of=pro2] {Update $\rho$ using $p^{(n+1)}$
and $T^{(n+1)}$};
\node (pro4) [process, below of=pro3] {Update $\vartheta_f$};
\node (dec1) [decision, below of=pro4, yshift=-0.4cm]
{Eq.~(\ref{eq:nonlinConv}) satisfied?}; 
\draw [arrow] (pro0) -- (pro1);
\draw [arrow] (pro1) -- (pro2);
\draw [arrow] (pro2) -- (pro3);
\draw [arrow] (pro3) -- (pro4);
\draw [arrow] (pro4) -- (dec1);
\draw [arrow] (dec1) --+(-2.7cm,0) |- (pro1);
\draw [arrow] (dec1) --+(+2.7cm,0) |- (pro0);
\node at (-2.2,-7.7) {no};
\node at (2.2,-7.7) {yes};
\node [rotate=90] at (-2.9,-4.7) {$n \leftarrow n+1$};
\node [rotate=90] at (2.9,-4.7) {$t \leftarrow t+\Delta t$};
\end{tikzpicture} \label{fig:singleloop}}
\qquad
\subfloat[Dual-loop solution procedure]
{\begin{tikzpicture}[node distance=1.5cm]
\node (pro0) [process] {Update previous time-levels};
\node (pro1) [process, below of=pro0] {Assemble and solve 
Eq.~(\ref{eq:eqsys})}; 
\node (pro30) [process, below of=pro1] {Update $\rho$ using $p^{(n+1)}$
and $T^{(m)}$};
\node (pro3) [process, below of=pro30] {Update $\vartheta_f$};
\node (dec1) [decision, below of=pro3, yshift=-0.4cm]
{Eq.~(\ref{eq:nonlinConv}) satisfied?}; 
\node (pro4) [process, below of=dec1, yshift=-0.6cm] {Update $T^{(m+1)}$ from
$h^{(m+1)}$}; \node (pro5) [process, below of=pro4] {Update $\rho$ using
$p^{(n+1)}$ and $T^{(m+1)}$}; 
\node (dec2) [decision, below of=pro5, yshift=-0.4cm]
{Eq.~(\ref{eq:nonlinDensity}) satisfied?}; 
\draw [arrow] (pro0) -- (pro1);
\draw [arrow] (pro1) -- (pro30);
\draw [arrow] (pro30) -- (pro3);
\draw [arrow] (pro3) -- (dec1);
\draw [arrow] (dec1) -- (pro4);
\draw [arrow] (pro4) -- (pro5);
\draw [arrow] (pro5) -- (dec2);
\draw [arrow] (dec1) --+(-2.7cm,0) |- (pro1);
\draw [arrow] (dec2) --+(-3.8cm,0) |- (pro1);
\draw [arrow] (dec2) --+(+2.7cm,0) |- (pro0);
\draw [densely dashed] (-3.3,-7.7) rectangle (2.2,-0.7);
\node at (-2.2,-6.2) {no};
\node at (-2.2,-11.7) {no};
\node at (2.2,-11.7) {yes};
\node at (0.4,-7.4) {yes};
\node at (-2.5,-7.5) {inner loop};
\node [rotate=90] at (-2.9,-4) {$n \leftarrow n+1$};
\node [rotate=90] at (2.9,-6.3) {$t \leftarrow t+\Delta t$};
\node [rotate=90] at (-4,-6.3) {$m \leftarrow m+1$};
\end{tikzpicture} \label{fig:dualloop}}
\end{small}
\caption{Flow charts of the a) single-loop and b) dual-loop solution
procedures. Note that the temperature $T$, which is only used to evaluate the
density $\rho$, is updated at different positions in the solution sequence,
requiring an additional nonlinear iteration loop for the dual-loop procedure.}
\label{fig:solution}
\end{center}
\end{figure}
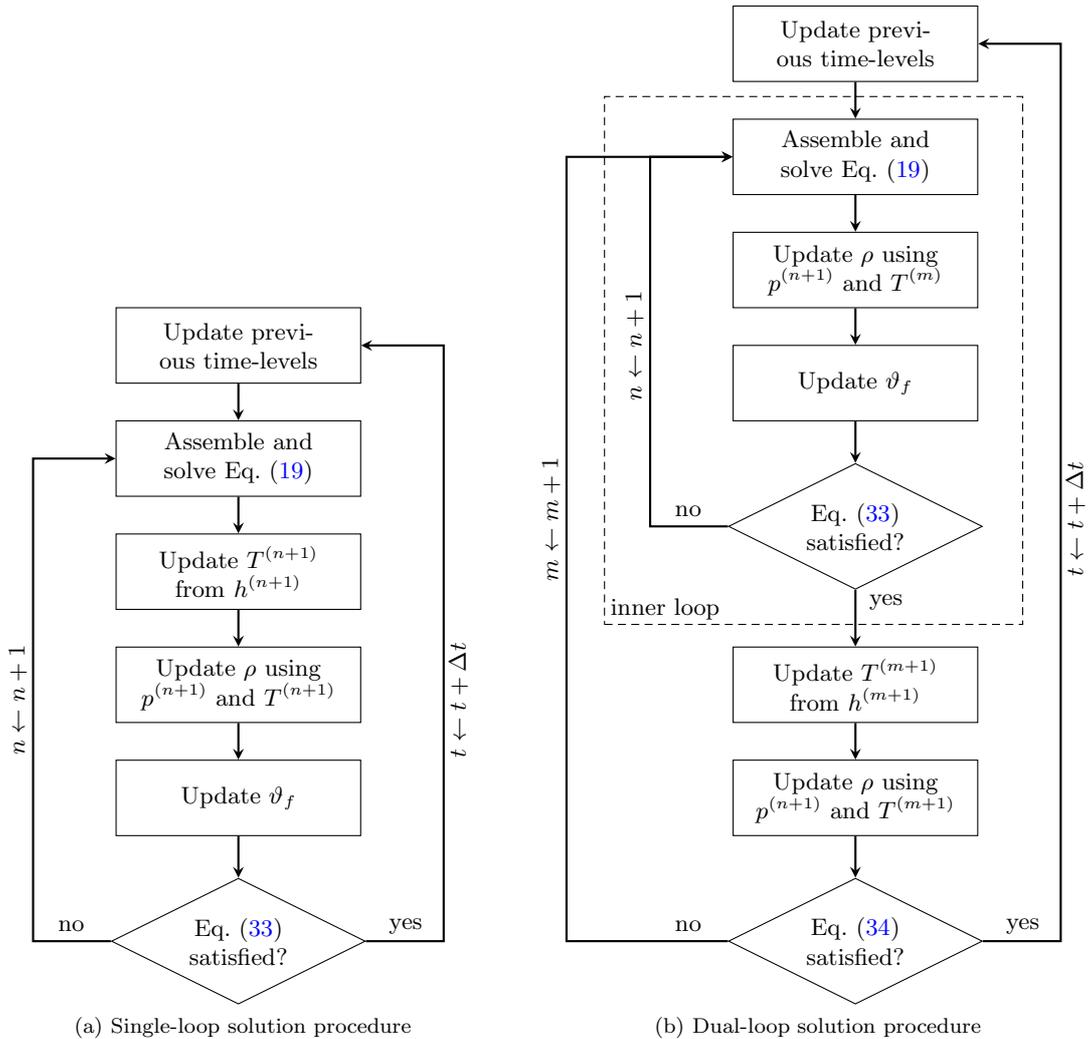

A {\em single-loop} and a {\em dual-loop} solution procedure are considered,
both schematically illustrated in Fig.~\ref{fig:solution}. The single-loop
solution procedure applies a straightforward update of all deferred (lagged)
variables after each nonlinear iteration. The dual-loop solution procedure is
based on the work of \citet{Xiao2017}, who proposed to introduce an {\em inner
loop}, in which the temperature used to update the density is assumed constant,
{\em i.e.}~the flow is assumed to be barotropic for the purpose of evaluating
the density and, hence, density is only a function of the pressure. Note that
the flow is not assumed to be isothermal in the inner loop, just the temperature
used to update the density is treated as constant. Once the nonlinear governing
equations in the inner loop have converged, the density is re-evaluated in an
{\em outer loop} based on the pressure and the updated temperature.
The density has converged if
\begin{equation}
\varepsilon_\rho^{(m)}=\sqrt{\frac{1}{N_\phi} \sum_{k=1}^{N_\phi}
\left( \frac{\phi_{\rho,k}^{(m+1)} -
\phi_{\rho,k}^{(m)}}{\phi_{\rho,k}^{(m)}}\right)^2} < \eta \ ,
\label{eq:nonlinDensity}
\end{equation}
where $\boldsymbol{\phi}_\rho$ is the vector of size $N_\phi$ that holds the
density $\rho$ at every cell centre of the computational mesh (hence, $N_\phi$
is equal to the number of mesh cells).
The dual-loop solution procedure is continued until both
Eq.~(\ref{eq:nonlinConv}) and Eq.~(\ref{eq:nonlinDensity}) are satisfied
simultaneously. This dual-loop solution procedure was shown to be stable for a
wide range of compressible flows in all Mach number regimes \cite{Xiao2017},
without the need for underrelaxation. It is noteworthy, however, that a
fixed-coefficient linearisation is applied in the algorithm of \citet{Xiao2017}
for the momentum equations and the energy equation.

\section{Results}
\label{sec:results}
With the aim of analysing the performance and stability associated with
different linearisation and solution strategies, four different test-cases are
considered: the propagation of acoustic waves in Section
\ref{sec:resultsAcoustic}, a shock tube in Section \ref{sec:shockTube}, as well
as the supersonic flow over a forward-facing step in Section \ref{sec:ffs} and
over a circular cone in Section \ref{sec:cone}. These test-cases cover Mach
numbers in the range $10^{-3} \lesssim M \leq 2$ as well as one- and
multi-dimensional simulations on Cartesian and tetrahedral meshes. Air is used
as the working fluid in all presented simulations, with $\gamma = 1.4$ and $c_v
= 720 \, \textup{J} \, \textup{kg}^{-1}  \, \textup{K}^{-1}$. The interested
reader is referred to the work of \citet{Xiao2017} for an extensive analysis of
the accuracy of the used numerical framework. In order to analyse the
convergence behaviour of the different linearisation and solution strategies,
the rate of convergence of the nonlinear equation system is estimated as
\begin{align}
q_n  & = \frac{\log \left( \| \boldsymbol{r}^{(n)} \| \right)}{\log
\left( \| \boldsymbol{r}^{(n+1)} \| \right)} \ , \label{eq:qn} \\
q_m  & = \frac{\log \left(\varepsilon_\rho^{(m)} \right)}{\log
\left(\varepsilon_\rho^{(m+1)} \right)} \label{eq:qm} \ .
\end{align}

\subsection{Propagation of acoustic waves}
\label{sec:resultsAcoustic}
The performance of simulations of the propagation of acoustics waves relies on a
robust coupling of density with pressure and temperature. At the same time, the
fluxes are small and momentum transport is an insignificant factor for the
stability and performance of the algorithm. The propagation of acoustic waves
is, thus, well suited to study the thermodynamic coupling of the solution
algorithm. 

\begin{figure}
\begin{center}
\includegraphics[width=0.47\textwidth]{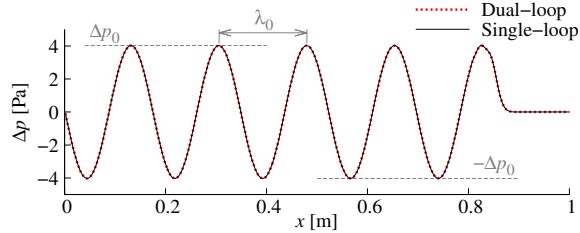}
\caption{Pressure profiles of the acoustic waves at $t=2.5 \times
10^{-3} \, \textup{s}$ using the single-loop and dual-loop solution procedures.
The theoretical pressure amplitude $\Delta p_0 = \pm \rho_0 a_0 \Delta u_0$ and
wavelength $\lambda_0 = a_0/f$ according to linear acoustic theory are given as
a reference.}
\label{fig:acoustic}
\end{center}
\end{figure}

The acoustic waves are simulated in a one-dimensional domain with mesh spacing
$\Delta x = 0.002 \, \textup{m}$, with a solution tolerance of $\eta =
10^{-12}$. The domain is initialised with a uniform pressure $p_0 = 10^5 \,
\textup{Pa}$, temperature $T_0 = 300 \, \textup{K}$ and velocity $u_0 = 1.0 \,
\textup{m} \, \textup{s}^{-1}$. The flow is perturbed by the velocity at the
domain-inlet, defined as $u_\textup{in} = u_0 + \Delta u \sin{(2 \pi f t)}$,
where $f=2000 \, \textup{s}^{-1}$ is the frequency and $\Delta u = 0.01 u_0$ is
the amplitude of the acoustic waves. Unless stated otherwise, the applied
time-step $\Delta t$ corresponds to a Courant number of $\textup{Co} = a_0
\Delta t/\Delta x = 0.1$, where $a_0=347.8 \, \textup{m} \, \textup{s}^{-1}$ is
the speed of sound according to Eq.~(\ref{eq:soundSpeed}).
The simulations are conducted on a single core of an Intel Xeon processor with
Haswell architecture. The pressure profiles for the acoustic waves in air using
the single-loop and the dual-loop solution procedures are shown in
Fig.~\ref{fig:acoustic}. The pressure amplitudes of the acoustic waves are in
excellent agreement with the theoretical pressure amplitude $\Delta p_0 = \pm
\rho_0 a_0 \Delta u$ based on linear acoustic theory \citep{Anderson2003}, and
the waves have the correct wavelength $\lambda_0 = a_0/f$. Furthermore, as
expected, no difference between the results obtained with either solution
procedure are observed for the acoustic waves.

The execution time $\tau$ for the simulation of these acoustic waves using
different linearisation and solution strategies are given in Table
\ref{tab:acousticSingleTime}. The Newton linearisation of the transient terms of
the momentum and energy equations yields a clear improvement in performance,
with a speedup of factor $1.4$ to $1.5$ compared to the fixed-coefficient
linearisation. Interestingly, while the simulations do not converge if the
single-loop solution procedure is applied in conjunction with a
fixed-coefficient linearisation of all terms, the single-loop solution procedure
converges, and yields a shorter execution time than the dual-loop solution
procedure, when the Newton linearisation is applied to the transient terms. The
linearisation of the advection terms of the governing equations, however, does
not have a significant impact on the execution times for the cases presented in
Table \ref{tab:acousticSingleTime}. In fact, this is to be expected considering
the small local changes in advecting velocity $\vartheta_f$ ({\em i.e.}~the
fluxes) and the low Mach number $M\approx 10^{-3}$, with the associated marginal
changes in density $\rho$. Increasing the time-step $\Delta t$, the additional
numerical stability associated with the $\vartheta$-Newton linearisation of the advection terms in
the momentum and energy equations becomes apparent, see Table
\ref{tab:acousticAdvCo}. If the fixed-coefficient linearisation is applied to
the advection terms, the execution time of the simulation increases
significantly as the Courant number exceeds unity and the solution algorithm
fails to converge for $\textup{Co}=10$. However, applying the $\vartheta$-Newton
linearisation, convergence is stable and rapid for all tested Courant numbers.
Note that the amplitude and the wavelength of the acoustic waves are not
predicted accurately for $\textup{Co} > 1$, with the amplitude decaying and the
wavelength increasing as the waves propagate downstream.

\begin{table}
\begin{center}
\caption{Execution time $\tau$ for the propagation of the
acoustic waves with different linearisation and solution
strategies.}
\label{tab:acousticSingleTime}
\begin{tabular}{llllrr}
\multirow{2}{*}{Case} & Continuity &
 \multicolumn{2}{c}{Momentum and energy} &
 \multicolumn{2}{c}{$\tau$ [s]}  \\
 & Advection & Transient & Advection & Dual-loop &  Single-loop \\
\hline
A & fixed-coeff. & fixed-coeff. & fixed-coeff. & $2529$ & -- \\
B & Newton & fixed-coeff. & fixed-coeff. & $2521$ & -- \\
C & fixed-coeff. & Newton & fixed-coeff. & $1668$ & $1211$ \\
D & Newton & Newton & fixed-coeff. &  $1697$ & $1192$ \\
E & Newton & Newton & $\rho$-Newton & $1731$ & $1231$ \\
F & Newton & Newton & $\vartheta$-Newton & $1718$ & $1254$ \\
G & Newton & Newton & full-Newton & $1786$ & $1211$ 
\end{tabular}
\end{center}
\end{table}

\begin{table}
\begin{center}
\caption{Execution time $\tau$ for the propagation of the acoustic waves,
simulated with different Courant numbers $\textup{Co}$ and different
linearisation strategies applied to the advection terms of the momentum and
energy equations, using the single-loop solution procedure. The Newton
linearisation is applied to the advection term of the continuity equation and the transient terms of the
momentum and energy equations.}
\label{tab:acousticAdvCo}
\begin{tabular}{lrrrr}
\multirow{2}{*}{Linearisation} &
\multicolumn{4}{c}{$\tau$ [s]} \\
 & $\textup{Co} = 0.5$ & $\textup{Co} = 1$ & $\textup{Co} = 2$ & $\textup{Co}
 = 10$ \\
\hline 
fixed-coeff. & $316$ & $248$ & $527$
& -- \\
  $\vartheta$-Newton  & $325$ & $191$ & $124$ &
$43$  \\
\end{tabular}
\end{center}
\end{table}

In summary, in this low Mach number case, the Newton linearisation of the
transient terms in the momentum and energy equations provides a significant
speedup, whereas the linearisation of the advection terms does not have a
sizeable impact on the performance of the numerical algorithm, since changes in
fluxes and density are small. However, at large Courant numbers the
$\vartheta$-Newton linearisation provides an improved stability and convergence
of the solution algorithm.

\subsection{Shock tube}
\label{sec:shockTube}
Due to their conceptual simplicity and well-defined theoretical solution, shock
tubes are frequently used test-cases for the validation and comparison of
numerical methods. The considered shock tube, which was originally proposed by
\citet{Sod1978}, features a shock wave, a rarefaction fan and a contact
discontinuity and, hence, provides a comprehensive test-case to analyse the
performance and convergence behaviour of the considered linearisation and solution
strategies.

\begin{figure}
\begin{center}
\subfloat[Density $\rho$]{
\includegraphics[width=0.32\textwidth]{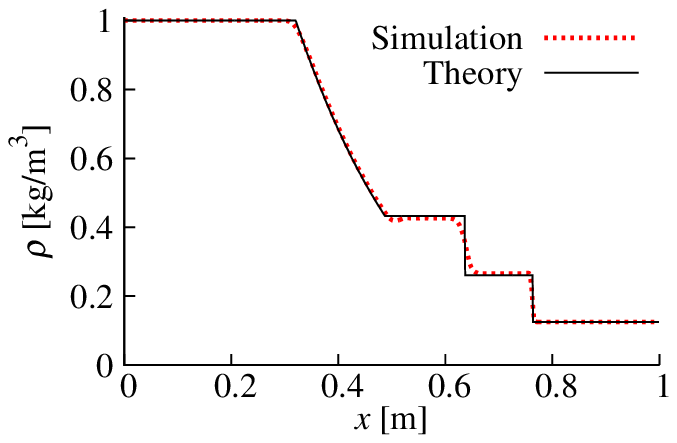}}
\qquad
\subfloat[Pressure $p$]{
\includegraphics[width=0.32\textwidth]{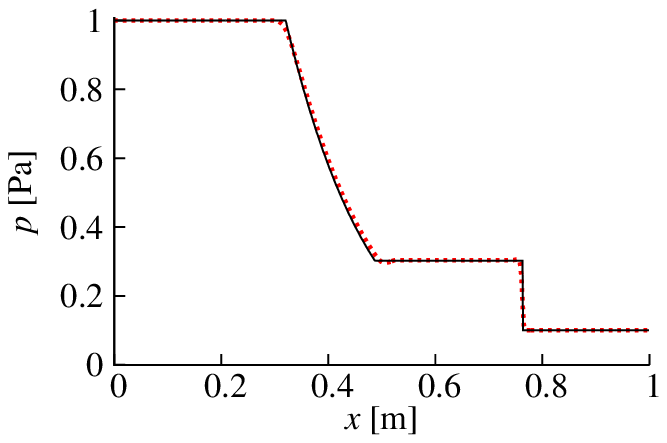}}
\caption{Density and pressure profiles of the shock tube at
$t=0.15 \, \textup{s}$, obtained with $\textup{Co} = 0.1$. The theoretical
Riemann solution is shown as a reference.}
\label{fig:sod0p1}
\end{center}
\end{figure}

\begin{figure}
\begin{center}
\subfloat[Density $\rho$]{
\includegraphics[width=0.32\textwidth]{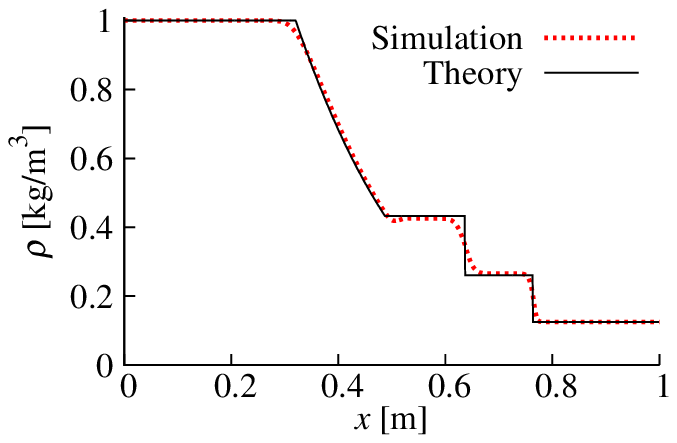}}
\qquad
\subfloat[Pressure $p$]{
\includegraphics[width=0.32\textwidth]{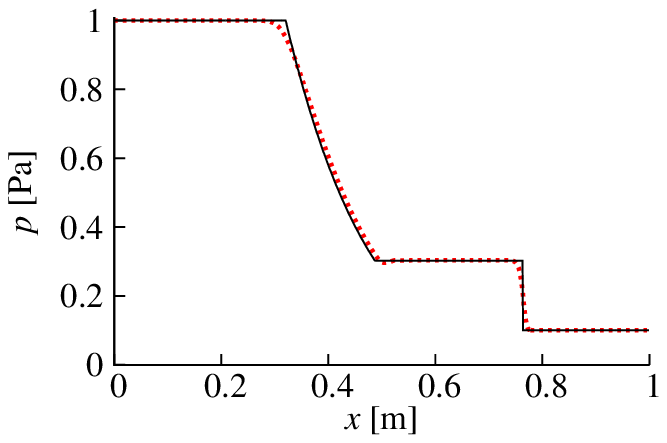}}
\caption{Density and pressure profiles of the shock tube at
$t=0.15 \, \textup{s}$, obtained with $\textup{Co} = 0.5$. The theoretical
Riemann solution is shown as a reference.}
\label{fig:sod0p5}
\end{center}
\end{figure}

The discontinuity of initial conditions separating the left state and the right
state is initially located in the middle of the one-dimensional domain with a
length of $1 \, \textup{m}$, which is represented with $400$ equidistant cells.
The initial conditions of the left and right states are
\citep{Sod1978}
\begin{equation}
\begin{array}{lllll}
u_\textup{L} = 0 \, \textup{m} \, \textup{s}^{-1}, & p_\textup{L} = 1.0 \,
\textup{Pa}, & \rho_\textup{L} = 1.000 \, \textup{kg} \, \textup{m}^{-3},
\nonumber \\
u_\textup{R} = 0 \, \textup{m} \, \textup{s}^{-1}, & p_\textup{R} = 0.1 \,
\textup{Pa}, & \rho_\textup{R} = 0.125 \, \textup{kg} \, \textup{m}^{-3}.
\nonumber
\end{array}
\end{equation}
The applied time-steps $\Delta t$ correspond to $\textup{Co} = a_\textup{L}
\Delta t/ \Delta x \in \{ 0.1,0.5 \}$ and the applied solution tolerance is
$\eta = 10^{-8}$. The simulations are conducted on a single core of an Intel
Xeon processor with Haswell architecture. The results for all considered
linearisation and solution strategies are in very good agreement with each other
and the theoretical Riemann solution for both considered Courant numbers, as
seen in Figs.~\ref{fig:sod0p1} and \ref{fig:sod0p5}.


The execution times $\tau$ for $\textup{Co} = 0.1$, listed in Table
\ref{tab:sod}, exhibit a similar pattern as observed for the propagation of acoustic waves in
Section \ref{sec:resultsAcoustic}. Using the dual-loop solution procedure, the
Newton linearisation of the transient terms of the momentum and energy equations
provides an appreciable speedup. The Newton linearisation of the transient terms
also yields converged solutions if the single-loop solution procedure is
applied, resulting in further speedup. The Newton linearisation of the
advection terms, on the other hand, has no significant impact on the performance
of the solution algorithm. Increasing the Courant number to $\textup{Co} = 0.5$,
for which the execution times are also given in Table \ref{tab:sod}, the dual-loop solution
procedure does not yield a converged result if all transient and advection terms are
linearised with the fixed-coefficient linearisation, whereas the Newton
linearisation of the advection term of the continuity equation exhibits a clear
performance benefit. In addition, with the single-loop solution
procedure, it is necessary to apply the $\vartheta$-Newton linearisation to the
advection terms of the momentum and energy equations to yield a converged
solution.

\begin{table}
\begin{center}
\caption{Execution time $\tau$ for the shock tube with
$\textup{Co} \in \{0.1,0.5\}$, using different linearisation and solution
strategies.}
\label{tab:sod}
\begin{tabular}{llllrrrr}
\multirow{2}{*}{Case} & Continuity & \multicolumn{2}{c}{Momentum and energy} &
\multicolumn{2}{c}{$\tau$ [s] for $\textup{Co}=0.1$} & \multicolumn{2}{c}{$\tau$
[s] for $\textup{Co}=0.5$}  \\
 & Advection & Transient & Advection & Dual-loop & Single-loop & Dual-loop &
Single-loop
\\
\hline
A & fixed-coeff. & fixed-coeff. & fixed-coeff. & $437$ & -- &
-- & -- \\
B & Newton & fixed-coeff. & fixed-coeff. & $413$ & -- &
$137$ & -- \\
C & fixed-coeff. & Newton & fixed-coeff. & $325$ & $142$ & 
$235$ & -- \\
D & Newton & Newton & fixed-coeff. & $315$ & $144$ & 
$111$ & -- \\
E & Newton & Newton & $\rho$-Newton &  $299$ & $137$ & 
$111$ & -- \\
F & Newton & Newton & $\vartheta$-Newton & $317$ & $150$ & 
$103$ & $49$ \\
G & Newton & Newton & full-Newton & $299$ & $152$ & 
$98$ & $41$
\end{tabular}
\end{center}
\end{table}

\begin{figure}
\begin{center}
\subfloat[Case A]{
\includegraphics[width=0.32\textwidth]{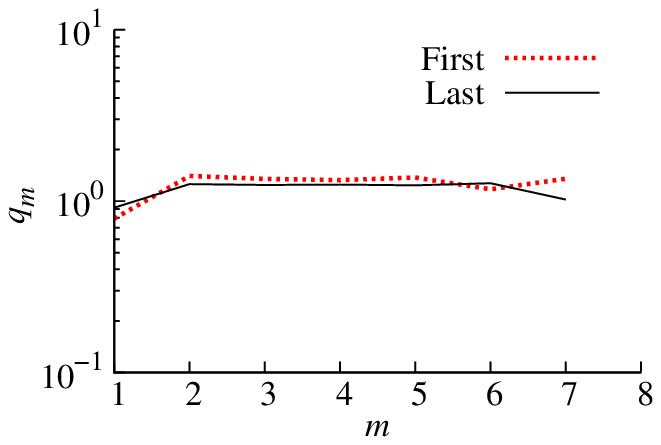}\label{fig:sodConvDoubleMA}}
\
\subfloat[Case C]{
\includegraphics[width=0.32\textwidth]{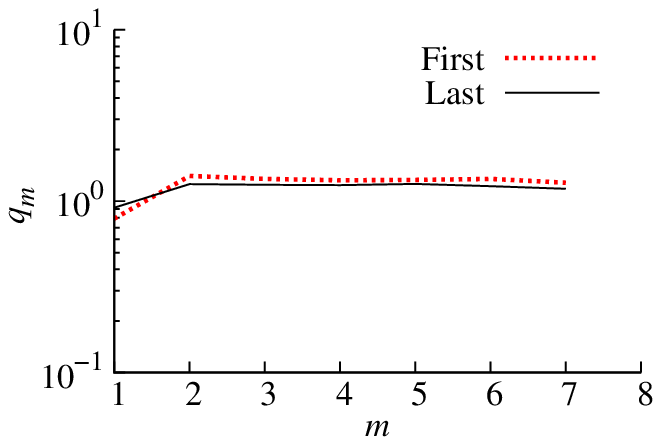}\label{fig:sodConvDoubleMB}}
\
\subfloat[Case G]{
\includegraphics[width=0.32\textwidth]{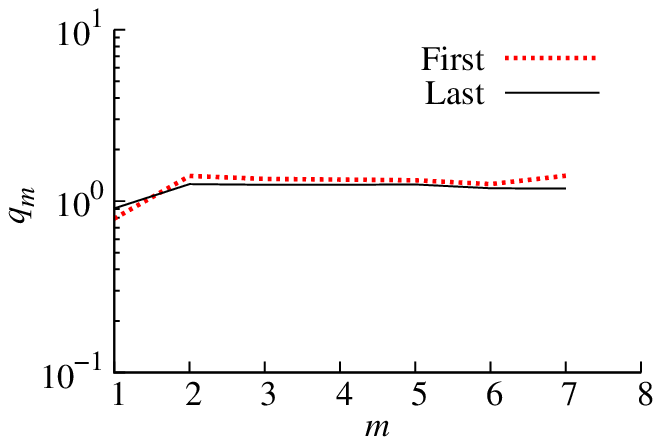}\label{fig:sodConvDoubleMC}}
\caption{Rate of convergence $q_m$, Eq.~(\ref{eq:qm}), for the shock tube,
obtained with the dual-loop solution procedure and $\textup{Co} = 0.1$, of the
first and last time-steps for different linearisation strategies (see Table
\ref{tab:sod}).}
\label{fig:sodConvDoubleM}
\end{center}
\end{figure}

\begin{figure}
\begin{center}
\subfloat[Case A]{\includegraphics[width=0.32\textwidth]{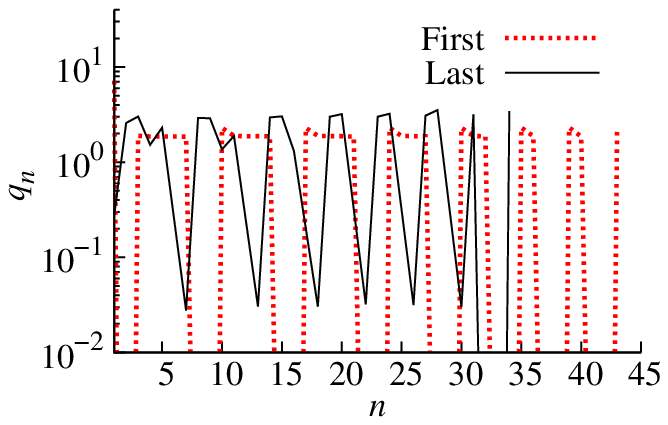}
\label{fig:sodConvDoubleNA}} \
\subfloat[Case C]{\includegraphics[width=0.32\textwidth]{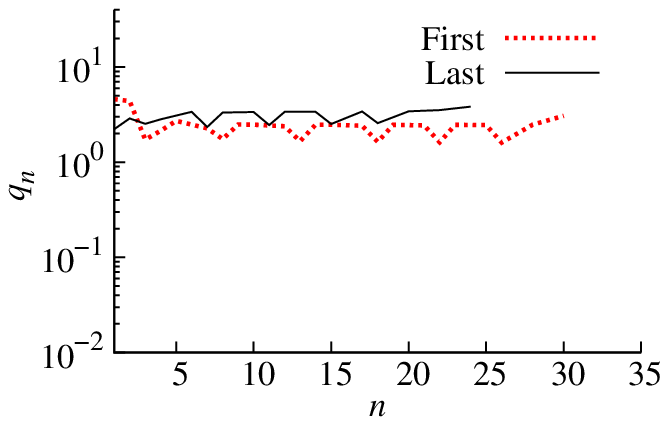}
\label{fig:sodConvDoubleNB}} \
\subfloat[Case G]{\includegraphics[width=0.32\textwidth]{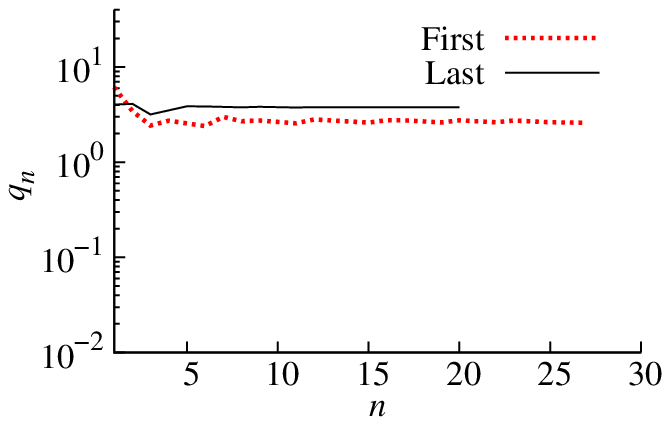}
\label{fig:sodConvDoubleNC}} 
\caption{Rate of convergence $q_n$, Eq.~(\ref{eq:qn}), for the shock
tube, obtained with the dual-loop solution procedure and
$\textup{Co} = 0.1$, of the first and last time-steps for different linearisation strategies
(see Table \ref{tab:sod}).}
\label{fig:sodConvDoubleN}
\end{center}
\end{figure}

\begin{figure}
\begin{center}
\subfloat[Case D]{
\includegraphics[width=0.32\textwidth]{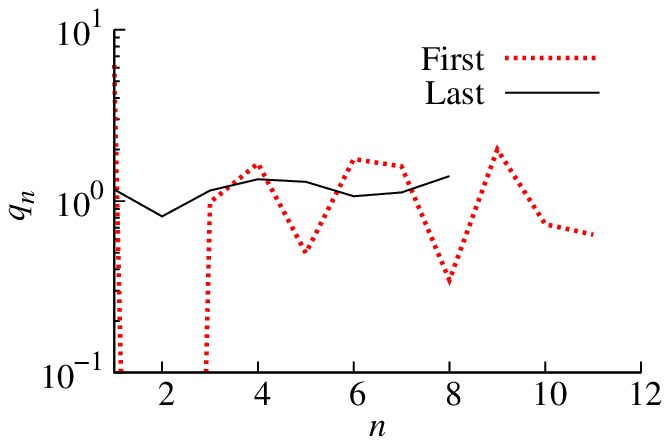}}
\
\subfloat[Case E]{
\includegraphics[width=0.32\textwidth]{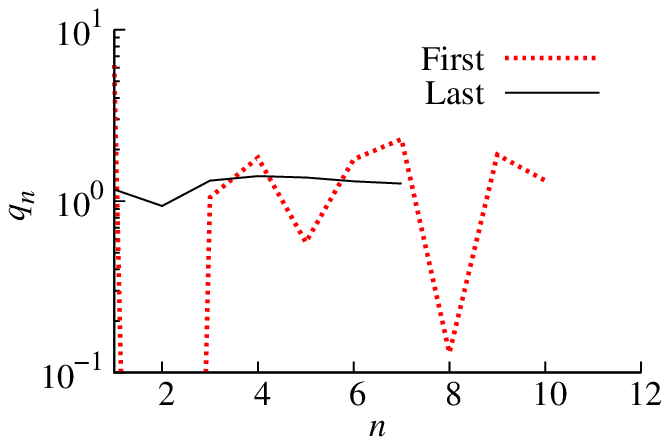}}
\
\subfloat[Case F]{
\includegraphics[width=0.32\textwidth]{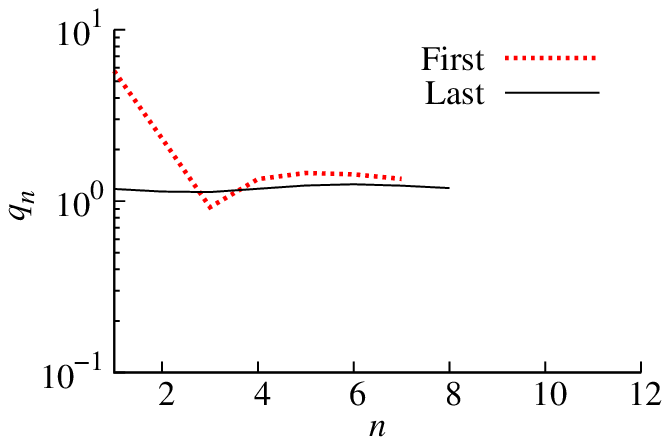}}
\caption{Rate of convergence $q_n$, Eq.~(\ref{eq:qn}), for the shock
tube, obtained with the single-loop solution procedure and
$\textup{Co} = 0.1$, of the first and last time-steps for different linearisation strategies
(see Table \ref{tab:sod}).}
\label{fig:sodConvSingle}
\end{center}
\end{figure}

The convergence rates of the outer loop $q_m$ and the inner loop $q_n$ for the
first and last time-steps of the simulations conducted with the dual-loop
solution procedure and $\textup{Co} = 0.1$ are shown in
Figs.~\ref{fig:sodConvDoubleM} and \ref{fig:sodConvDoubleN}, respectively, for
three different linearisation strategies. The outer loop converges with a
similar and almost constant convergence rate of $q_m \approx 1.25$ in all three
cases, as seen Fig.~\ref{fig:sodConvDoubleM}. However, large differences in the
convergence behaviour can be observed in Fig.~\ref{fig:sodConvDoubleN} for the
inner loop.
Case A, which corresponds to a fixed-coefficient linearisation for all nonlinear
terms, exhibits strong oscillations of the convergence rate $q_n$, see
Fig.~\ref{fig:sodConvDoubleNA}; in the first time-step $q_n$ even becomes
negative after each outer loop. Applying a Newton linearisation to the transient
terms of the momentum and energy equations (Case C), see
Fig.~\ref{fig:sodConvDoubleNB}, reduces the amplitude of these oscillations of
the convergence rate $q_n$ substantially, circumventing negative convergence
rates. The convergence becomes even smoother when a Newton linearisation is
applied to all nonlinear terms (Case G), with $q_n \approx 2.6$ in the first
time-step and $q_n \approx 3.8$ in the last time-step, as seen in
Fig.~\ref{fig:sodConvDoubleNC}. Examining the convergence obtained with the
single-loop solution procedure, shown in Fig.~\ref{fig:sodConvSingle}, shows
that the full-Newton linearisation of the advection terms of the momentum and
energy equations (Case G) yields a smooth convergence behaviour, while applying
only the $\rho$-Newton (Case E) or the $\vartheta$-Newton (Case F)
linearisations yields oscillations of the convergence rate $q_n$. The
convergence rate $q_n$ is nominally lower with the single-loop solution
procedure than with the dual-loop procedure, which is attributed to the stronger
nonlinearity of the governing equations, because density is dependent on both
pressure and temperature simultaneously using the single-loop solution
procedure.


\subsection{Supersonic flow over a forward-facing step}
\label{sec:ffs}
\begin{figure}
\begin{center}
\subfloat[Mach number $M$]
{\includegraphics[width=0.47\textwidth]{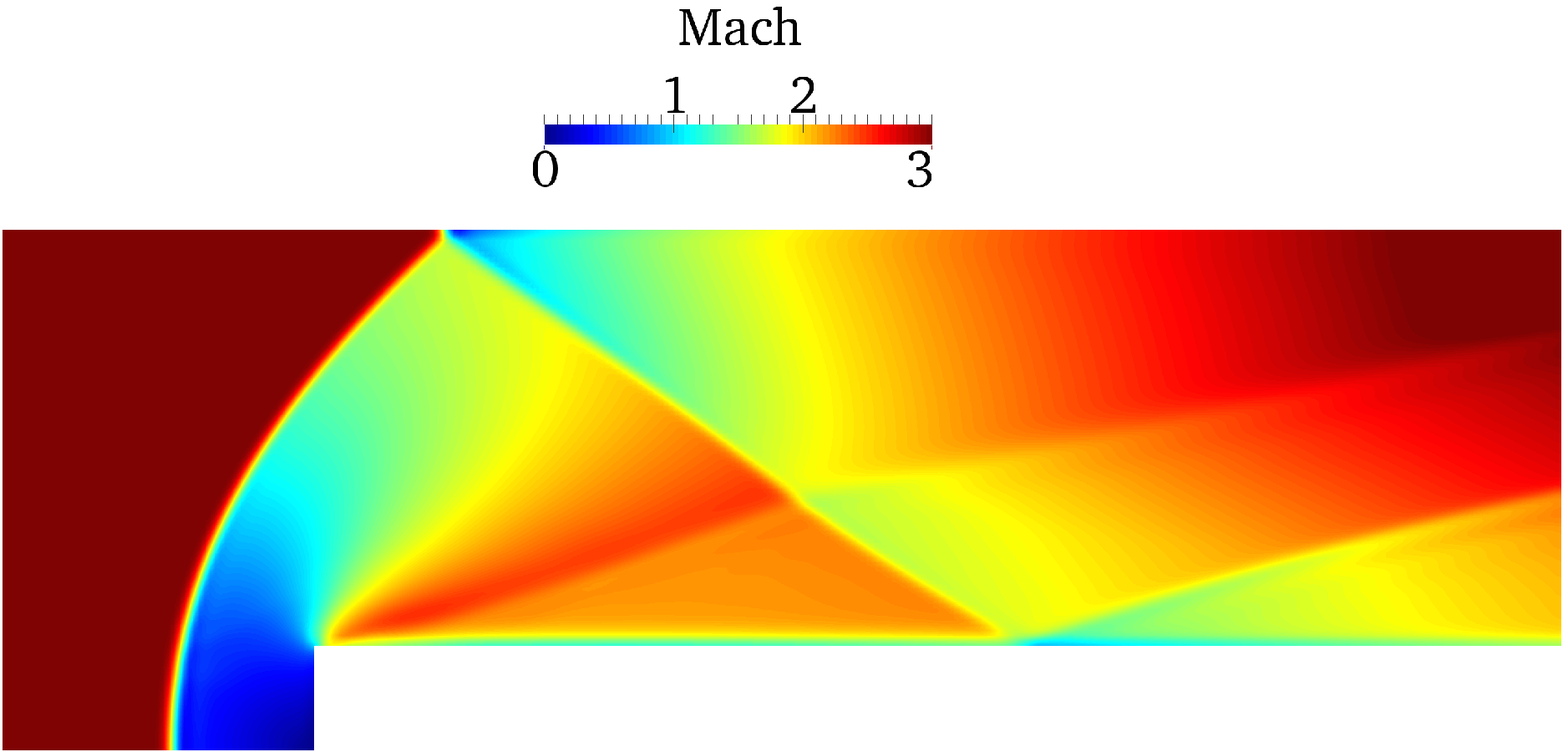}}
\quad
\subfloat[Pressure $p$]
{\includegraphics[width=0.47\textwidth]{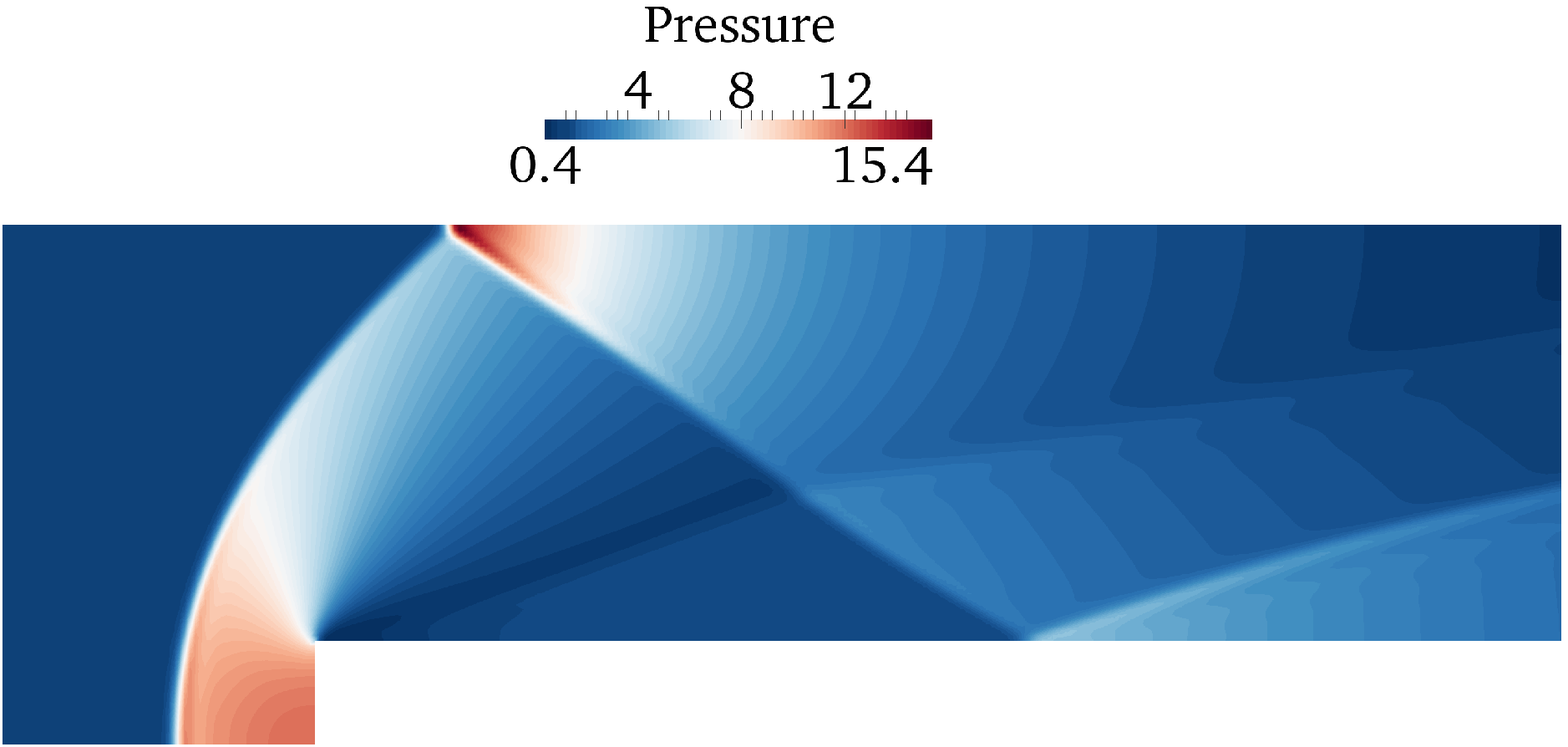}}
\caption{Mach number and pressure contours of the supersonic flow over a
forward-facing step at $t=2 \, \textup{s}$, with $\textup{Co} = 0.9$.}
\label{fig:ffs2}
\subfloat[Mach number $M$]
{\includegraphics[width=0.47\textwidth]{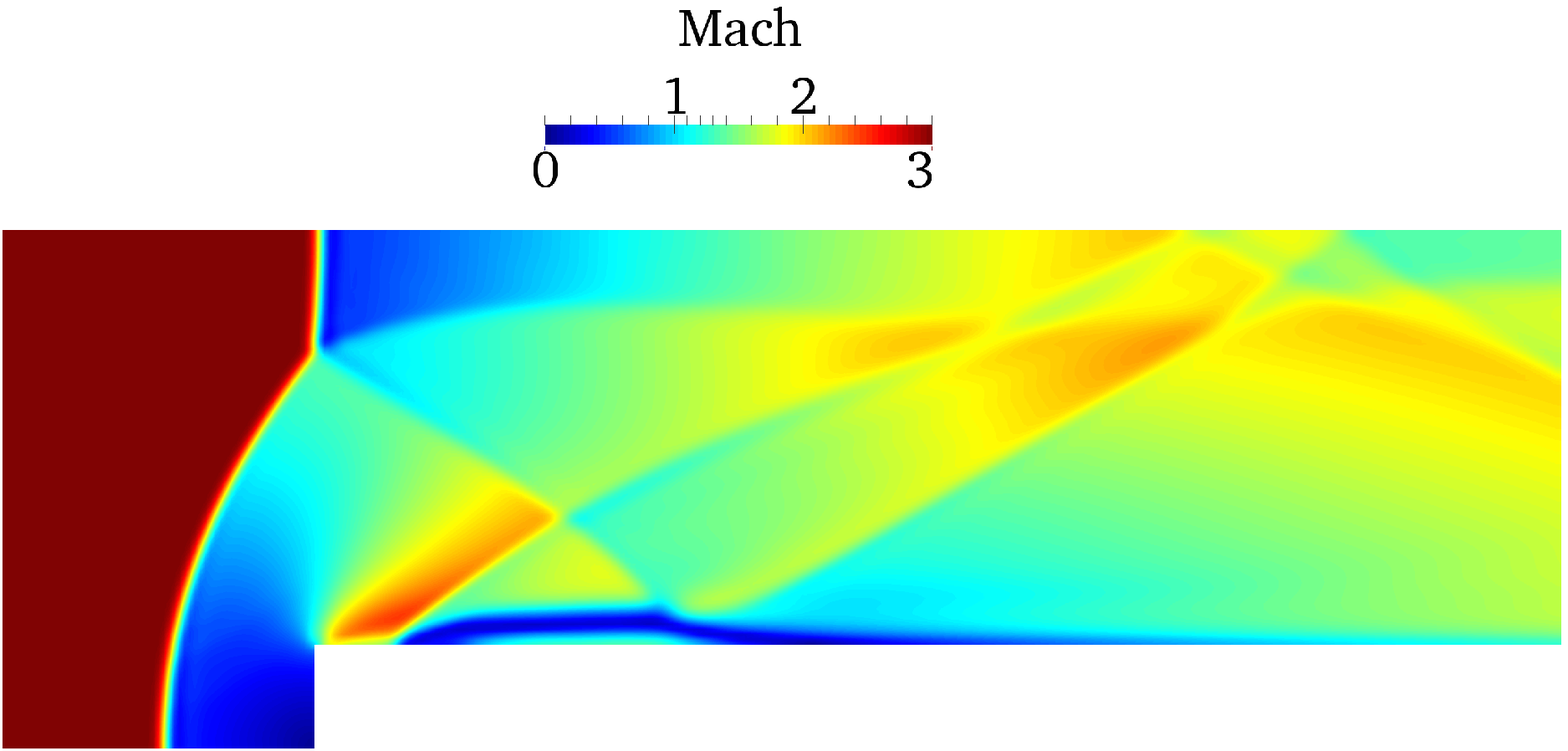}}
\quad
\subfloat[Pressure $p$]
{\includegraphics[width=0.47\textwidth]{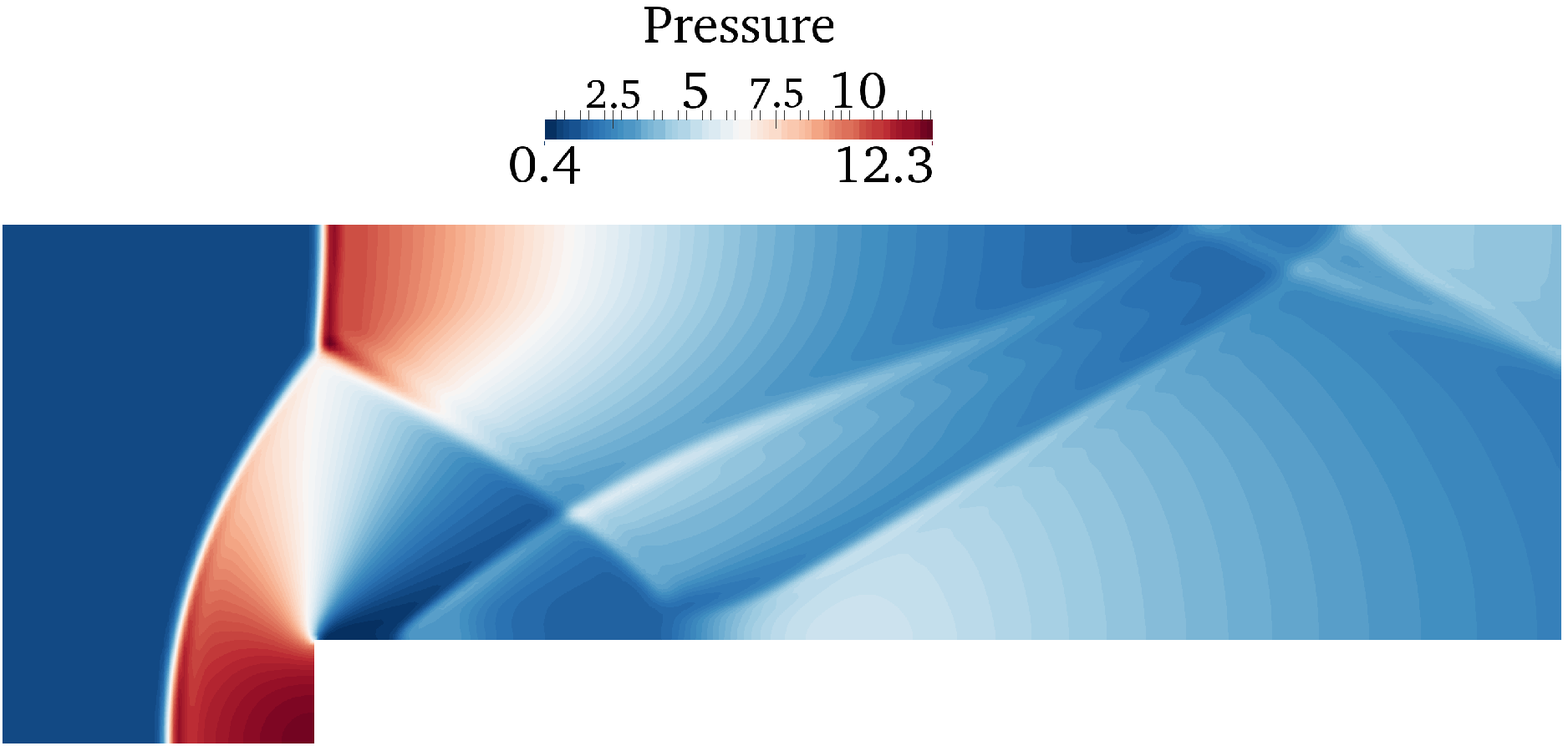}}
\caption{Mach number and pressure contours of the supersonic flow over a
forward-facing step at $t=4 \, \textup{s}$, with $\textup{Co} = 0.9$.}
\label{fig:ffs4}
\end{center}
\end{figure}

The two-dimensional supersonic flow over a forward-facing step is frequently
used to test new numerical methods and algorithms. Following \citet{Woodward1984}, the
computational domain is $3 \, \textup{m} \times 1 \, \textup{m}$ with a step of
height $0.2 \, \textup{m}$, positioned at $x=0.6 \, \textup{m}$. The flow
entering the domain has a Mach number of $M = u/a_0 = 3$. The mesh spacing of
the applied equidistant Cartesian mesh is $\Delta x = 0.01 \, \textup{m}$, the
applied time-steps $\Delta t$ correspond to $\textup{Co} = u \Delta t/ \Delta x
\in \{0.3,0.9\}$, and the applied solution tolerance is $\eta = 10^{-7}$.  The
particular challenge of this test-case is the spatiotemporally 
evolving shock waves and the associated development of a transonic flow, as well
as large pressure gradients.
Figures \ref{fig:ffs2} and \ref{fig:ffs4} show the Mach number and pressure
contours of the evolving transonic flow at $t=2\, \textup{s}$ and $t=4\,
\textup{s}$, respectively, which are in good agreement with previously reported
results \cite{Woodward1984, Jasak1996}. The simulations are conducted on a
single compute node equipped with two Intel Xeon processors (Haswell
architecture) containing $10$ cores each.

\begin{figure}
\begin{center}
\subfloat[Case B]{
\includegraphics[width=0.32\textwidth]{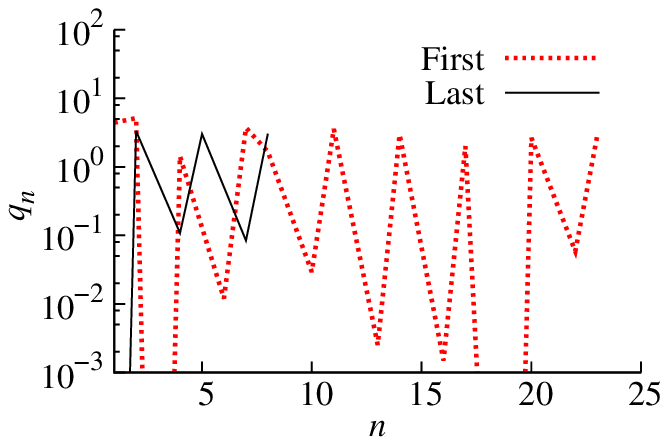}} \
\subfloat[Case D]{
\includegraphics[width=0.32\textwidth]{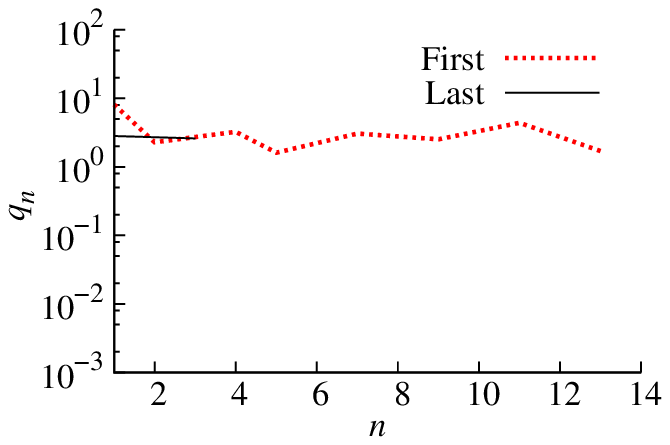}} \
\subfloat[Case G]{
\includegraphics[width=0.32\textwidth]{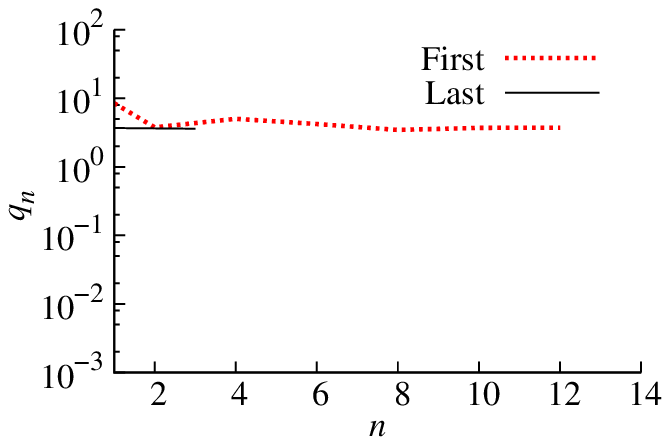}}
\caption{Rate of convergence $q_n$, Eq.~(\ref{eq:qn}), for the flow
over a forward-facing step, obtained with the dual-loop solution procedure and
$\textup{Co} = 0.3$, of the first and last time-steps for different
linearisation strategies (see Table \ref{tab:ffs}).}
\label{fig:ffsConvDualN}
\end{center}
\end{figure}

\begin{figure}
\begin{center}
\subfloat[Case E]{
\includegraphics[width=0.32\textwidth]{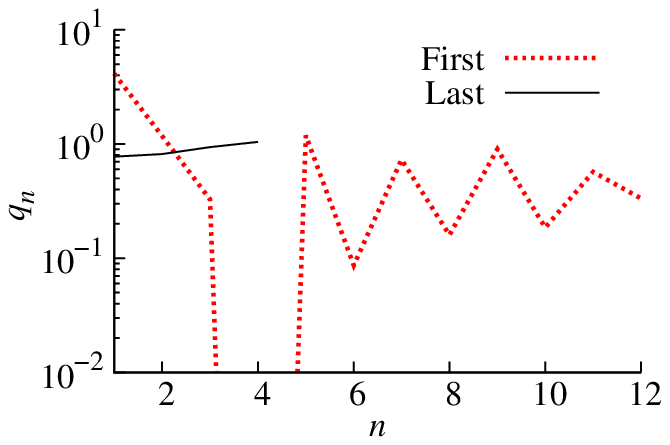}} \
\subfloat[Case G]{
\includegraphics[width=0.32\textwidth]{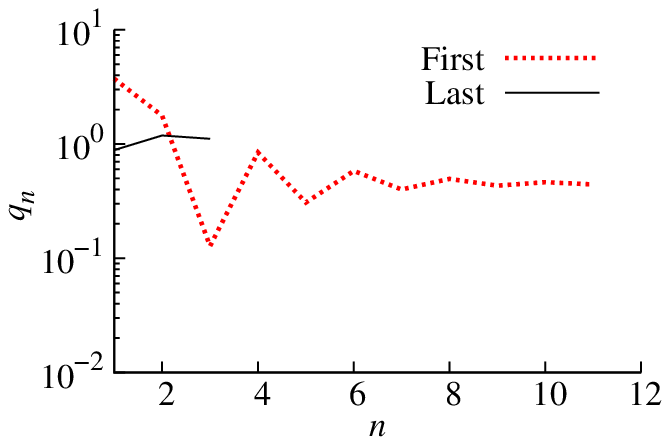}}
\caption{Rate of convergence $q_n$, Eq.~(\ref{eq:qn}), for the flow
over a forward-facing step, obtained with the single-loop solution procedure and
$\textup{Co} = 0.9$, of the first and last time-steps for different
linearisation strategies (see Table \ref{tab:ffs}).}
\label{fig:ffsConvSingle}
\end{center}
\end{figure}

The execution times $\tau$ for $\textup{Co} = 0.3$ are given in Table
\ref{tab:ffs}, using both the single- and dual-loop solution procedures. Note
that applying a fixed-coefficient linearisation to the advection term of the
continuity equation does not yield a converged solution for the considered
Courant numbers with either of the applied solution procedures. In conjunction
with the dual-loop solution procedure, the reduction in execution time as a
result of applying the Newton linearisation to the transient terms of the
momentum and energy equations, instead of the fixed-coefficient linearisation,
is similar to the cases discussed in the previous sections.
Although the linearisation of the advection terms has no substantial impact with
respect to the solution time, the convergence rate $q_n$ of the inner loop is
less oscillatory applying a Newton linearisation to the transient or advection
terms, as seen in Fig.~\ref{fig:ffsConvDualN}.
If the time-step is increased to $\textup{Co}=0.9$, the $\rho$-Newton
linearisation of the advection terms in the momentum and energy equations turns
out to be crucial with respect to the performance and stability of the solution
algorithm, as seen in Table \ref{tab:ffs}. In fact, a converged result is
obtained only with the $\rho$-Newton linearisation when the single-loop solution
procedure is applied.
The fully implicit treatment of density, through the Newton linearisation of the
transient terms together with the $\rho$-Newton linearisation of the advection
terms, provides a strong implicit pressure-density coupling, which is
particularly significant in the transonic flow regime.
Nevertheless, applying the full-Newton linearisation by adding the
$\vartheta$-Newton linearisation further improves the convergence behaviour and
circumvents negative convergence rates, as seen in Fig.~\ref{fig:ffsConvSingle},
albeit with only a small reduction of the execution time.

\begin{table}
\begin{center}
\caption{Execution time $\tau$ for the flow over a
forward-facing step with $\textup{Co} \in \{0.3,0.9\}$, using different
linearisation and solution strategies.}
\label{tab:ffs}
\begin{tabular}{llllrrrr}
\multirow{2}{*}{Case} & Continuity & \multicolumn{2}{c}{Momentum and energy} &
\multicolumn{2}{c}{$\tau \, [\textup{s}]$ for $\textup{Co}=0.3$} &
\multicolumn{2}{c}{$\tau \, [\textup{s}]$ for $\textup{Co}=0.9$}  \\
 & Advection & Transient & Advection & Dual-loop & Single-loop & Dual-loop &
Single-loop \\
\hline
B & Newton & fixed-coeff. & fixed-coeff. & $4966$ & -- &
$5846$ & -- \\
D & Newton & Newton & fixed-coeff. & $2689$ & $1818$ &
$2886$ & -- \\
E & Newton & Newton & $\rho$-Newton & $2763$ & $1841$ &
$1769$ & $1035$ \\
F & Newton & Newton & $\vartheta$-Newton & $2967$ & $1925$ &
$3079$ & -- \\
G & Newton & Newton & full-Newton & $2860$ & $1940$ &
$1711$ & $921$
\end{tabular}
\end{center}
\end{table}


\subsection{Supersonic flow over a cone}
\label{sec:cone}
\begin{figure}[t]
\begin{center}
\subfloat[Schematic illustration]{
\includegraphics[width=0.33\textwidth]{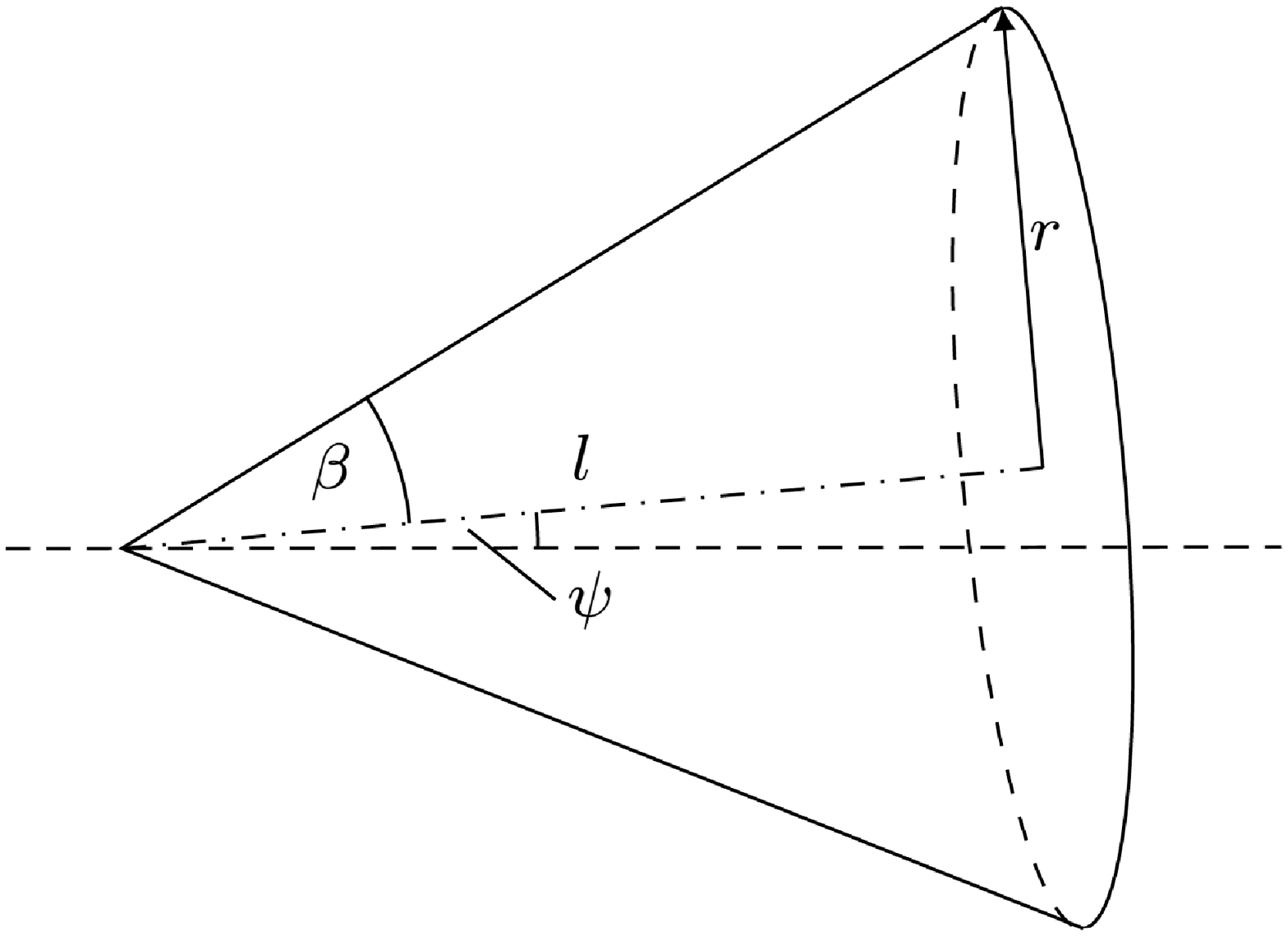}
\label{fig:coneSchematic}} \qquad
\subfloat[Mach number contours with computational mesh]
{\includegraphics[width=0.4\textwidth]{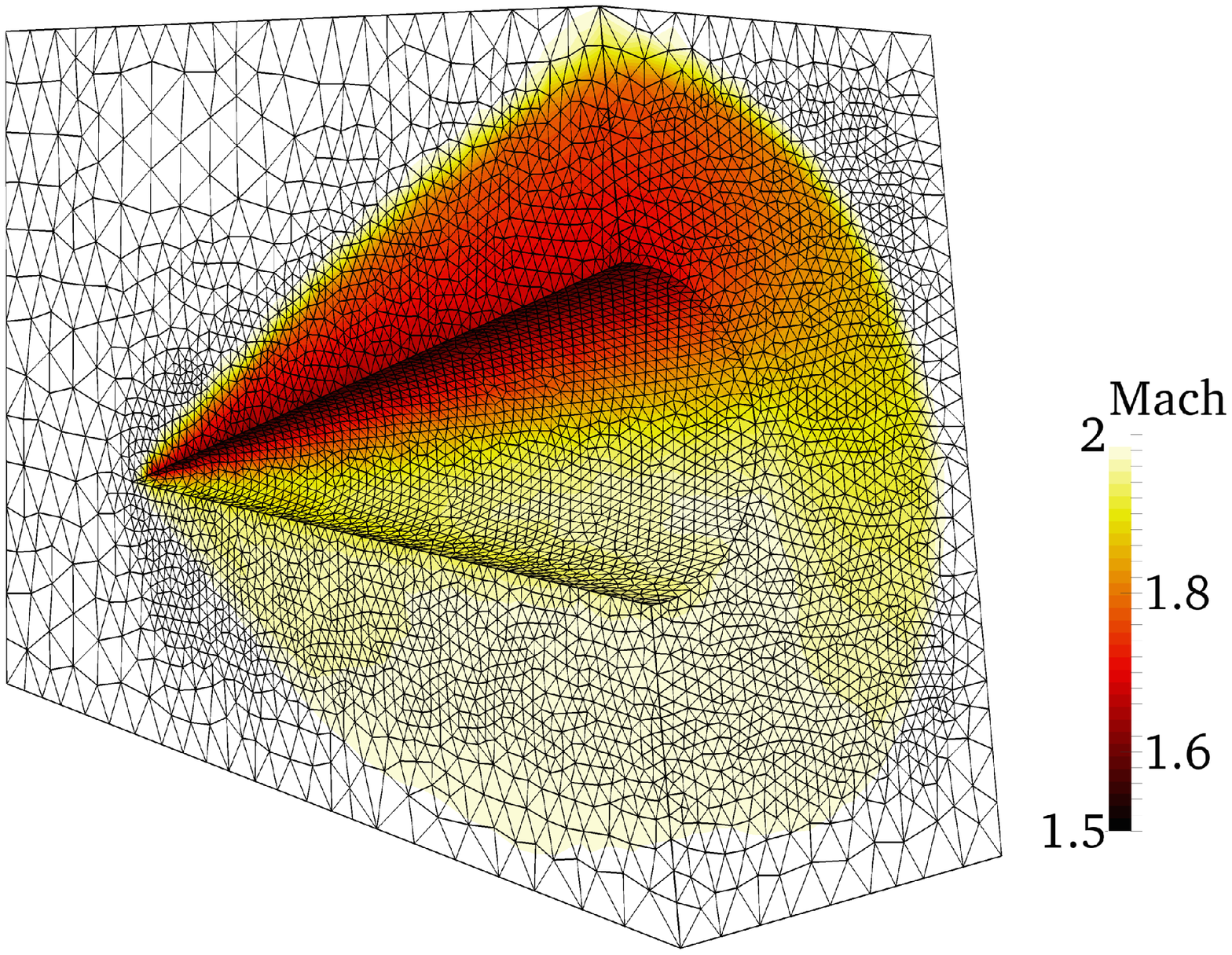}
\label{fig:coneMesh}}
 \caption{Schematic illustration of the circular cone with
 radius $r$, length $l$, cone angle $\beta$ and angle of attack $\psi$ and
 steady-state Mach number contours with the applied computational mesh of the
 flow with $M=2$ over the considered circular cone ($r=0.05 \, \textup{m}$, $l =
 0.1 \, \textup{m}$, $\beta = 10^\circ$, $\psi = 10^\circ$).}
 \label{fig:cone}
\end{center}
\end{figure}

As a final test-case, the three-dimensional supersonic flow over a circular cone
is simulated. The cone, schematically shown in Fig.~\ref{fig:coneSchematic}, has
a radius of $r=0.05 \, \textup{m}$, a length of $l = 0.1 \, \textup{m}$ and the
cone angle is $\beta = 10^\circ$. The flow with $M = 2$ is oriented with an
angle of attack of $\psi = 10^\circ$ to the primary axis of the cone. Because of
the symmetry of the flow, only half of the cone is simulated, in a computational
domain represented by a tetrahedral mesh with approximately $7.41 \times 10^5$
cells, shown in Fig.~\ref{fig:coneMesh} together with the Mach number contours
at steady state.
The applied time-step corresponds to $\textup{Co} = 0.54$ and the solution
tolerance is $\eta = 10^{-7}$. Following \citet{Xiao2017}, the domain is
initialised with uniform pressure $p_0 = 10^5 \, \textup{Pa}$, temperature $T_0
= 300 \, \textup{K}$ and velocity $u_0 = 695.59 \, \textup{m} \,
\textup{s}^{-1}$, corresponding to $M=2$. \citet{Xiao2017} compared the results
obtained with the applied numerical framework for supersonic flows over
different circular cones favourably against previous studies
\citep{Sims1964,Kutler1971}.
The presented simulations are stopped at $t=7.5 \times 10^{-5} \, \textup{s}$,
at which point the flow has assumed a steady state. The simulations are
conducted on a single compute node equipped with two Intel Xeon processors
(Haswell architecture) containing $10$ cores each.

\begin{figure}
\begin{center}
\subfloat[Dual-loop solution procedure]{
\includegraphics[width=0.32\textwidth]{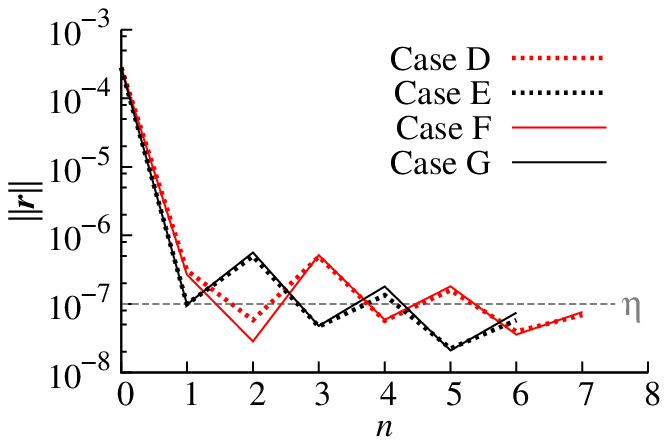}} \
\subfloat[Single-loop solution procedure]{
\includegraphics[width=0.32\textwidth]{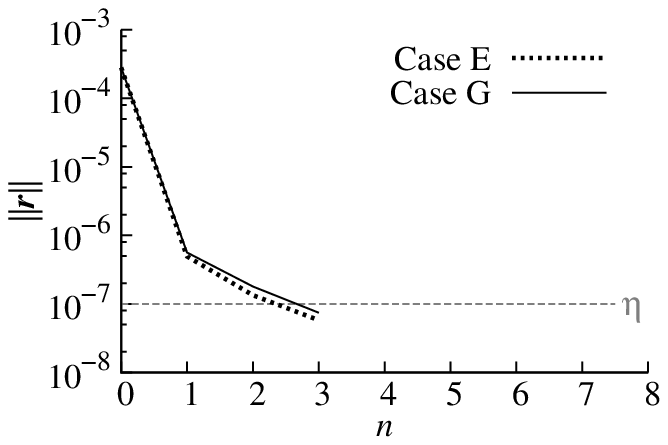}}
\caption{$L_2$-norm of the residual vector $\boldsymbol{r}$,
Eq.~(\ref{eq:nonlinConv}), of the first-time step for the supersonic flow over a
circular cone, using a) the dual-loop solution procedure and b) the single-loop solution
procedure, with different linearisation strategies. Black (red) lines are used
for cases with (without) $\rho$-Newton linearisation of the advection terms.
Note that for the shown cases using the dual-loop solution procedure, every
maximum in $\| \boldsymbol{r} \|$ is associated with an increment of the
outer loop, $m \leftarrow m+1$, see Fig.~\ref{fig:dualloop}, {\em i.e.}~it
follows a density update based on pressure $p$ and the updated temperature $T$.}
\label{fig:coneR}
\end{center}
\end{figure}

The execution times $\tau$ of the simulation with both the single- and dual-loop
solution procedures are given in Table \ref{tab:cone}.
Similar to the flow over the forward-facing step in Section \ref{sec:ffs},
simulations without Newton linearisation of the advection term of the continuity
equation do not yield a converged solution for the considered Courant number.
In addition, even for the dual-loop solution procedure, a Newton linearisation
of the transient terms of the momentum and energy equations is required for
convergence. The $\rho$-Newton linearisation of the advection terms in the
momentum and energy equations is found to be critical for the performance and
stability of the solution algorithm, as similarly observed in Section
\ref{sec:ffs}, in particular using the single-loop solution procedure. The
difference in execution time between the dual-loop and the single-loop solution
procedures is noticeably smaller than in all other considered cases, which may
be attributed to the strong coupling of pressure and density in the supersonic
regime. In particular, with the Newton linearisation of the transient terms and
the $\rho$-Newton linearisation of the advection terms, pressure and density are
coupled implicitly in both the single-loop and the dual-loop solution
procedures; the implicit coupling of the equation system, thus, closely
represents the nature of the flow. At the same time, the influence of changes
of the fluxes, {\em i.e.}~the advecting velocity $\vartheta_f$, are less
significant in the supersonic regime, which explains the small impact of the
$\vartheta$-Newton linearisation of the advection terms. This can also be
observed in Fig.~\ref{fig:coneR}, which shows the residual norms obtained with
both solution procedures; a clear difference in convergence behaviour can be
seen between cases with and without $\rho$-Newton linearisation, whereas the
$\vartheta$-Newton linearisation has an almost negligible influence on the
convergence.

\begin{table}
\begin{center}
\caption{Execution time $\tau$ for the supersonic flow over a
cone with $M = 2$ and $\textup{Co}=0.54$, using different linearisation and
solution strategies.}
\label{tab:cone}
\begin{tabular}{llllrr}
\multirow{2}{*}{Case} & Continuity & \multicolumn{2}{c}{Momentum and energy} &
\multicolumn{2}{c}{$\tau \, [\textup{s}]$}  \\
 & Advection & Transient & Advection & Dual-loop & Single-loop \\
\hline
B & Newton & fixed-coeff. & fixed-coeff. & -- & -- \\
D & Newton & Newton & fixed-coeff. & $19517$ & -- \\
E & Newton & Newton & $\rho$-Newton & $12117$ & $10616$ \\
F & Newton & Newton & $\vartheta$-Newton & $17130$ & -- \\
G & Newton & Newton & full-Newton & $12137$ & $10485$
\end{tabular}
\end{center}
\end{table}

\section{Conclusions}
\label{sec:conclusions}
Different linearisation and iterative solution strategies have been analysed and
compared in the context of a fully-coupled pressure-based algorithm for
compressible flows at all speeds, with the aim of elucidating their impact on
performance and stability of the algorithm. To this end, the analysis has
focused on test-cases with compression and expansion waves in all Mach
number regimes. The presented results highlight a substantial influence
of the chosen linearisation and provide new insight into the design of
efficient and robust pressure-based algorithms for compressible flows. The discussed
single-loop and dual-loop solution algorithms do not feature underrelaxation
procedures or other tuning parameters, and are, therefore, straightforward in
their application; although the reduction of nonlinearity through a
barotropic density update in the inner loop of the dual-loop solution procedure
is perhaps somewhat akin to an underrelaxation. 

The strong implicit coupling of pressure, density and velocity through a Newton
linearisation of the transient terms of the momentum and energy equations was
found to be the primary performance driver in all Mach number regimes, providing
a speedup of up to factor $2.2$ for the considered test-cases. The
linearisation of the transient term was further observed to be a prerequisite
for the application of the single-loop solution procedure, resulting in a further
reduction of the execution time for flows in all Mach number regimes. The reason
for this improved performance and stability is attributed to the smoother and less oscillatory
convergence behaviour of the iterative solution algorithm, in particular with
respect to the nonlinear residual. Even though, applying the dual-loop solution
procedure, the peak convergence rates of the inner loop are similarly high for
the considered test-cases, the convergence rate quickly drops below $1$ without
the Newton linearisation of the transient terms. 

The Newton linearisation of the advection terms of the continuity, momentum and
energy equations was found to have a negligible influence on the performance and
stability at low Mach numbers, in conjunction with low Courant numbers. In fact,
due to the increase in the number of non-zero coefficients of the sparse
coefficient matrix of the linear system of governing equations, the
$\vartheta$-Newton linearisation of the advection terms was found to
slightly increase the execution time for low Mach number flows, {\em e.g.}~the
propagation of acoustic waves. However, the Newton linearisation of the advection
term of the continuity equation becomes essential for the stability of the
solution algorithm for high Mach number flows. With regards to the linearisation
of the advection terms of the momentum and energy equations, the $\rho$-Newton
linearisation was also found to be important for the performance and stability
for flows with large Mach numbers. The $\vartheta$-Newton linearisation of the
advection terms improves the convergence and stability for flows in all Mach
number regimes when the Courant number is large, with both the single-loop and
dual-loop solution procedures. In fact, the only linearisation strategy that
yields a stable convergence for all considered test-cases, irrespective of the
considered Mach number, Courant number and solution strategy, is the full-Newton
linearisation of all advection terms in conjunction with the Newton
linearisation of the transient terms.

The dual-loop solution procedure has been shown to be, in general, more stable
than the single-loop solution procedure, owing to the reduction in nonlinearity
through the barotropic density update in the inner loop. This surplus in
stability comes at the cost of longer execution times. Hence, when the
single-loop solution procedure converges, it yields a significant reduction in execution
time; for the considered shock-tube, for instance, switching from the dual-loop
to the single-loop solution procedure reduces the execution time by factor $2.3$.

In summary, the presented study highlights the importance of a careful
linearisation of the governing nonlinear equations for compressible flows. To
this end, a full Newton linearisation of all transient and advection terms of the governing 
equations is found to be overall beneficial, improving the performance and
stability of the solution algorithm, and fully exploits the implicit coupling
via the simultaneous solution of the governing equations in the applied
fully-coupled pressure-based algorithm. The accelerated convergence and improved
stability of the solution algorithm, in conjunction with the
elimination of all underrelaxation measures and increase in the applied Courant
number, has been shown to speedup the simulations by several times compared to a
simple but widely applied fixed-coefficient linearisation, for flows in all
Mach number regimes.

\begin{samepage}
\section*{Acknowledgements}
\noindent  The author gratefully acknowledges financial support from the
Engineering and Physical Sciences Research Council (EPSRC) through grant EP/M021556/1. 
\end{samepage}


\begin{thebibliography}{42}
\expandafter\ifx\csname natexlab\endcsname\relax\def\natexlab#1{#1}\fi
\providecommand{\bibinfo}[2]{#2}
\ifx\xfnm\relax \def\xfnm[#1]{\unskip,\space#1}\fi
\bibitem[{Tannehill et~al.(1997)Tannehill, Pletcher, and
  Anderson}]{Anderson1997}
\bibinfo{author}{J.~Tannehill}, \bibinfo{author}{R.~Pletcher},
  \bibinfo{author}{D.~Anderson}, \bibinfo{title}{Computational {{Fluid
  Mechanics}} and {{Heat Transfer}}}, \bibinfo{publisher}{{Taylor \& Francis}},
  \bibinfo{edition}{second} edition, \bibinfo{year}{1997}.
\bibitem[{Wesseling(2001)}]{Wesseling2001}
\bibinfo{author}{P.~Wesseling}, \bibinfo{title}{Principles of {{Computational
  Fluid Dynamics}}}, \bibinfo{publisher}{{Springer}}, \bibinfo{year}{2001}.
\bibitem[{Beam and Warming(1978)}]{Beam1978}
\bibinfo{author}{R.~M. Beam}, \bibinfo{author}{R.~F. Warming},
\newblock \bibinfo{title}{An {{Implicit Factored Scheme}} for the
  {{Compressible Navier}}-{{Stokes Equations}}},
\newblock \bibinfo{journal}{AIAA Journal} \bibinfo{volume}{16}
  (\bibinfo{year}{1978}) \bibinfo{pages}{393--402}.
\bibitem[{MacCormack(1982)}]{MacCormack1982}
\bibinfo{author}{R.~W. MacCormack},
\newblock \bibinfo{title}{A {{Numerical Method}} for {{Solving}} the
  {{Equations}} of {{Compressible Viscous Flow}}},
\newblock \bibinfo{journal}{AIAA Journal} \bibinfo{volume}{20}
  (\bibinfo{year}{1982}) \bibinfo{pages}{1275--1281}.
\bibitem[{Turkel et~al.(1997)Turkel, Radespiel, and Kroll}]{Turkel1997}
\bibinfo{author}{E.~Turkel}, \bibinfo{author}{R.~Radespiel},
  \bibinfo{author}{N.~Kroll},
\newblock \bibinfo{title}{Assessment of preconditioning methods for
  multidimensional aerodynamics},
\newblock \bibinfo{journal}{Computers \& Fluids} \bibinfo{volume}{26}
  (\bibinfo{year}{1997}) \bibinfo{pages}{613--634}.
\bibitem[{{van der Heul} et~al.(2003){van der Heul}, Vuik, and
  Wesseling}]{vanderHeul2003}
\bibinfo{author}{D.~{van der Heul}}, \bibinfo{author}{C.~Vuik},
  \bibinfo{author}{P.~Wesseling},
\newblock \bibinfo{title}{A conservative pressure-correction method for flow at
  all speeds},
\newblock \bibinfo{journal}{Computers \& Fluids} \bibinfo{volume}{32}
  (\bibinfo{year}{2003}) \bibinfo{pages}{1113--1132}.
\bibitem[{Cordier et~al.(2012)Cordier, Degond, and Kumbaro}]{Cordier2012}
\bibinfo{author}{F.~Cordier}, \bibinfo{author}{P.~Degond},
  \bibinfo{author}{A.~Kumbaro},
\newblock \bibinfo{title}{An {{Asymptotic}}-{{Preserving}} all-speed scheme for
  the {{Euler}} and {{Navier}}\textendash{{Stokes}} equations},
\newblock \bibinfo{journal}{Journal of Computational Physics}
  \bibinfo{volume}{231} (\bibinfo{year}{2012}) \bibinfo{pages}{5685--5704}.
\bibitem[{Miettinen and Siikonen(2015)}]{Miettinen2015}
\bibinfo{author}{A.~Miettinen}, \bibinfo{author}{T.~Siikonen},
\newblock \bibinfo{title}{Application of pressure- and density-based methods
  for different flow speeds: {{Application}} of pressure- and density-based
  methods for different flow speeds},
\newblock \bibinfo{journal}{International Journal for Numerical Methods in
  Fluids} \bibinfo{volume}{79} (\bibinfo{year}{2015})
  \bibinfo{pages}{243--267}.
\bibitem[{Harlow and Amsden(1971)}]{Harlow1971a}
\bibinfo{author}{F.~H. Harlow}, \bibinfo{author}{A.~A. Amsden},
\newblock \bibinfo{title}{A numerical fluid dynamics calculation method for all
  flow speeds},
\newblock \bibinfo{journal}{Journal of Computational Physics}
  \bibinfo{volume}{8} (\bibinfo{year}{1971}) \bibinfo{pages}{197--213}.
\bibitem[{Van~Doormaal et~al.(1987)Van~Doormaal, Raithby, and
  McDonald}]{VanDoormaal1987}
\bibinfo{author}{J.~Van~Doormaal}, \bibinfo{author}{G.~Raithby},
  \bibinfo{author}{B.~McDonald},
\newblock \bibinfo{title}{The {{Segregated Approach}} to {{Predicting Viscous
  Compressible Fluid Flows}}},
\newblock \bibinfo{journal}{ASME Journal of Turbomachinery}
  \bibinfo{volume}{109} (\bibinfo{year}{1987}) \bibinfo{pages}{268--277}.
\bibitem[{Chen and Pletcher(1991)}]{Chen1991}
\bibinfo{author}{K.-H. Chen}, \bibinfo{author}{R.~Pletcher},
\newblock \bibinfo{title}{Primitive {{Variable}}, {{Strongly Implicit
  Calculation Procedure}} for {{Viscous Flows}} at {{All Speeds}}},
\newblock \bibinfo{journal}{AIAA Journal} \bibinfo{volume}{29}
  (\bibinfo{year}{1991}) \bibinfo{pages}{1241--1249}.
\bibitem[{Acharya et~al.(2007)Acharya, Baliga, Karki, Murthy, Prakash, and
  Vanka}]{Acharya2007}
\bibinfo{author}{S.~Acharya}, \bibinfo{author}{B.~R. Baliga},
  \bibinfo{author}{K.~Karki}, \bibinfo{author}{J.~Y. Murthy},
  \bibinfo{author}{C.~Prakash}, \bibinfo{author}{S.~P. Vanka},
\newblock \bibinfo{title}{Pressure-{{Based Finite}}-{{Volume Methods}} in
  {{Computational Fluid Dynamics}}},
\newblock \bibinfo{journal}{Journal of Heat Transfer} \bibinfo{volume}{129}
  (\bibinfo{year}{2007}) \bibinfo{pages}{407}.
\bibitem[{Moukalled et~al.(2016)Moukalled, Mangani, and
  Darwish}]{Moukalled2016}
\bibinfo{author}{F.~Moukalled}, \bibinfo{author}{L.~Mangani},
  \bibinfo{author}{M.~Darwish}, \bibinfo{title}{The Finite Volume Method in
  Computational Fluid Dynamics: {{An}} Advanced Introduction with {{OpenFOAM}}
  and {{Matlab}}}, \bibinfo{publisher}{{Springer}}, \bibinfo{year}{2016}.
\bibitem[{Harlow and Amsden(1968)}]{Harlow1968}
\bibinfo{author}{F.~H. Harlow}, \bibinfo{author}{A.~A. Amsden},
\newblock \bibinfo{title}{Numerical calculation of almost incompressible flow},
\newblock \bibinfo{journal}{Journal of Computational Physics}
  \bibinfo{volume}{3} (\bibinfo{year}{1968}) \bibinfo{pages}{80--93}.
\bibitem[{Issa et~al.(1986)Issa, Gosman, and Watkins}]{Issa1986}
\bibinfo{author}{R.~Issa}, \bibinfo{author}{A.~Gosman},
  \bibinfo{author}{A.~Watkins},
\newblock \bibinfo{title}{The computation of compressible and incompressible
  recirculating flows by a non-iterative implicit scheme},
\newblock \bibinfo{journal}{Journal of Computational Physics}
  \bibinfo{volume}{62} (\bibinfo{year}{1986}) \bibinfo{pages}{66--82}.
\bibitem[{Karki and Patankar(1989)}]{Karki1989}
\bibinfo{author}{K.~C. Karki}, \bibinfo{author}{S.~V. Patankar},
\newblock \bibinfo{title}{Pressure based calculation procedure for viscous
  flows at all speeds in arbitrary configurations},
\newblock \bibinfo{journal}{AIAA Journal} \bibinfo{volume}{27}
  (\bibinfo{year}{1989}) \bibinfo{pages}{1167--1174}.
\bibitem[{Issa and Javareshkian(1998)}]{Issa1998}
\bibinfo{author}{R.~I. Issa}, \bibinfo{author}{M.~H. Javareshkian},
\newblock \bibinfo{title}{Pressure-{{Based Compressible Calculation Method
  Utilizing Total Variation Diminishing Schemes}}},
\newblock \bibinfo{journal}{AIAA Journal} \bibinfo{volume}{36}
  (\bibinfo{year}{1998}) \bibinfo{pages}{1652--1657}.
\bibitem[{Moukalled and Darwish(2000)}]{Moukalled2000}
\bibinfo{author}{F.~Moukalled}, \bibinfo{author}{M.~Darwish},
\newblock \bibinfo{title}{A unified formulation of the segregated class of
  algorithms for fluid flow at all speeds},
\newblock \bibinfo{journal}{Numerical heat transfer, Part B.}
  \bibinfo{volume}{37} (\bibinfo{year}{2000}) \bibinfo{pages}{103--139}.
\bibitem[{Demir{\v d}zi{\'c} et~al.(1993)Demir{\v d}zi{\'c}, Lilek, and
  Peri{\'c}}]{Demirdzic1993}
\bibinfo{author}{I.~Demir{\v d}zi{\'c}}, \bibinfo{author}{v.~Lilek},
  \bibinfo{author}{M.~Peri{\'c}},
\newblock \bibinfo{title}{A collocated finite volume method for predicting
  flows at all speeds},
\newblock \bibinfo{journal}{International Journal for Numerical Methods in
  Fluids} \bibinfo{volume}{16} (\bibinfo{year}{1993})
  \bibinfo{pages}{1029--1050}.
\bibitem[{Karimian and Schneider(1994)}]{Karimian1994}
\bibinfo{author}{S.~M.~H. Karimian}, \bibinfo{author}{G.~E. Schneider},
\newblock \bibinfo{title}{Pressure-based computational method for compressible
  and incompressible flows},
\newblock \bibinfo{journal}{Journal of Thermophysics and Heat Transfer}
  \bibinfo{volume}{8} (\bibinfo{year}{1994}) \bibinfo{pages}{267--274}.
\bibitem[{Karimian and Schneider(1995)}]{Karimian1995}
\bibinfo{author}{S.~M.~H. Karimian}, \bibinfo{author}{G.~E. Schneider},
\newblock \bibinfo{title}{Pressure-based control-volume finite element method
  for flow at all speeds},
\newblock \bibinfo{journal}{AIAA Journal} \bibinfo{volume}{33}
  (\bibinfo{year}{1995}) \bibinfo{pages}{1611--1618}.
\bibitem[{Chen and Przekwas(2010)}]{Chen2010}
\bibinfo{author}{Z.~Chen}, \bibinfo{author}{A.~J. Przekwas},
\newblock \bibinfo{title}{A coupled pressure-based computational method for
  incompressible/compressible flows},
\newblock \bibinfo{journal}{Journal of Computational Physics}
  \bibinfo{volume}{229} (\bibinfo{year}{2010}) \bibinfo{pages}{9150--9165}.
\bibitem[{Darwish and Moukalled(2014)}]{Darwish2014}
\bibinfo{author}{M.~Darwish}, \bibinfo{author}{F.~Moukalled},
\newblock \bibinfo{title}{A fully coupled navier-stokes solver for fluid flow
  at all speeds},
\newblock \bibinfo{journal}{Numerical Heat Transfer, Part B: Fundamentals}
  \bibinfo{volume}{65} (\bibinfo{year}{2014}) \bibinfo{pages}{410--444}.
\bibitem[{Xiao et~al.(2017)Xiao, Denner, and {van Wachem}}]{Xiao2017}
\bibinfo{author}{C.-N. Xiao}, \bibinfo{author}{F.~Denner},
  \bibinfo{author}{B.~{van Wachem}},
\newblock \bibinfo{title}{Fully-coupled pressure-based finite-volume framework
  for the simulation of fluid flows at all speeds in complex geometries},
\newblock \bibinfo{journal}{Journal of Computational Physics}
  \bibinfo{volume}{346} (\bibinfo{year}{2017}) \bibinfo{pages}{91--130}.
\bibitem[{Kunz et~al.(1999)Kunz, Cope, and Venkateswaran}]{Kunz1999}
\bibinfo{author}{R.~Kunz}, \bibinfo{author}{W.~Cope},
  \bibinfo{author}{S.~Venkateswaran},
\newblock \bibinfo{title}{Development of an implicit method for multi-fluid
  flow simulations},
\newblock \bibinfo{journal}{Journal of Computational Physics}
  \bibinfo{volume}{152} (\bibinfo{year}{1999}) \bibinfo{pages}{78--101}.
\bibitem[{Dennis and Schnabel(1996)}]{Dennis1996}
\bibinfo{author}{J.~E. Dennis}, \bibinfo{author}{R.~B. Schnabel},
  \bibinfo{title}{Numerical {{Methods}} for {{Unconstrained Optimization}} and
  {{Nonlinear Equations}}}, \bibinfo{publisher}{{Society for Industrial and
  Applied Mathematics}}, \bibinfo{year}{1996}.
\bibitem[{Darbandi et~al.(2008)Darbandi, Roohi, and Mokarizadeh}]{Darbandi2008}
\bibinfo{author}{M.~Darbandi}, \bibinfo{author}{E.~Roohi},
  \bibinfo{author}{V.~Mokarizadeh},
\newblock \bibinfo{title}{Conceptual linearization of {{Euler}} governing
  equations to solve high speed compressible flow using a pressure-based
  method},
\newblock \bibinfo{journal}{Numerical Methods for Partial Differential
  Equations} \bibinfo{volume}{24} (\bibinfo{year}{2008})
  \bibinfo{pages}{583--604}.
\bibitem[{Darbandi and Mokarizadeh(2004)}]{Darbandi2004}
\bibinfo{author}{M.~Darbandi}, \bibinfo{author}{V.~Mokarizadeh},
\newblock \bibinfo{title}{A modified pressure-based algorithm to solve flow
  fields with shock and expansion waves},
\newblock \bibinfo{journal}{Numerical Heat Transfer, Part B: Fundamentals}
  \bibinfo{volume}{46} (\bibinfo{year}{2004}) \bibinfo{pages}{497--504}.
\bibitem[{Ferziger and Peri{\'c}(2002)}]{Ferziger2002}
\bibinfo{author}{J.~Ferziger}, \bibinfo{author}{M.~Peri{\'c}},
  \bibinfo{title}{Computational {{Methods}} for {{Fluid Dynamics}}},
  \bibinfo{publisher}{{Springer Verlag}}, \bibinfo{address}{Berlin Heidelberg
  New York}, \bibinfo{edition}{3.} edition, \bibinfo{year}{2002}.
\bibitem[{Roe(1986)}]{Roe1986}
\bibinfo{author}{P.~Roe},
\newblock \bibinfo{title}{Characteristic-based schemes for the euler
  equations},
\newblock \bibinfo{journal}{Annual Review of Fluid Mechanics}
  \bibinfo{volume}{18} (\bibinfo{year}{1986}) \bibinfo{pages}{337--365}.
\bibitem[{Denner et~al.(2018)Denner, Xiao, and {van Wachem}}]{Denner2018b}
\bibinfo{author}{F.~Denner}, \bibinfo{author}{C.-N. Xiao},
  \bibinfo{author}{B.~{van Wachem}},
\newblock \bibinfo{title}{Pressure-based algorithm for compressible interfacial
  flows with acoustically-conservative interface discretisation},
\newblock \bibinfo{journal}{Journal of Computational Physics}
  \bibinfo{volume}{367} (\bibinfo{year}{2018}) \bibinfo{pages}{192--234}.
\bibitem[{Demir{\v d}zi{\'c} and Muzaferija(1995)}]{Demirdzic1995}
\bibinfo{author}{I.~Demir{\v d}zi{\'c}}, \bibinfo{author}{S.~Muzaferija},
\newblock \bibinfo{title}{Numerical method for coupled fluid flow, heat
  transfer and stress analysis using unstructured moving meshes with cells of
  arbitrary topology},
\newblock \bibinfo{journal}{Computer Methods in Applied Mechanics and
  Engineering} \bibinfo{volume}{125} (\bibinfo{year}{1995})
  \bibinfo{pages}{235--255}.
\bibitem[{Balay et~al.(1997)Balay, Gropp, McInnes, and Smith}]{Balay1997}
\bibinfo{author}{S.~Balay}, \bibinfo{author}{W.~Gropp}, \bibinfo{author}{L.~C.
  McInnes}, \bibinfo{author}{B.~F. Smith},
\newblock \bibinfo{title}{Efficient {{Management}} of {{Parallelism}} in
  {{Object Oriented Numerical Software Libraries}}},
\newblock in: \bibinfo{editor}{E.~Arge}, \bibinfo{editor}{A.~Bruasat},
  \bibinfo{editor}{H.~Langtangen} (Eds.), \bibinfo{booktitle}{Modern {{Software
  Tools}} in {{Scientific Computing}}}, \bibinfo{publisher}{{Birkhaeuser
  Press}}, \bibinfo{year}{1997}, pp. \bibinfo{pages}{163--202}.
\bibitem[{Balay et~al.(2017{\natexlab{a}})Balay, Abhyankar, Adams, Brown,
  Brune, Buschelman, Dalcin, Eijkhout, Gropp, Kaushik, Knepley, McInnes, Rupp,
  Smith, Zampini, Zhang, and Zhang}]{petsc-web-page}
\bibinfo{author}{S.~Balay}, \bibinfo{author}{S.~Abhyankar},
  \bibinfo{author}{M.~F. Adams}, \bibinfo{author}{J.~Brown},
  \bibinfo{author}{P.~Brune}, \bibinfo{author}{K.~Buschelman},
  \bibinfo{author}{L.~Dalcin}, \bibinfo{author}{V.~Eijkhout},
  \bibinfo{author}{W.~D. Gropp}, \bibinfo{author}{D.~Kaushik},
  \bibinfo{author}{M.~G. Knepley}, \bibinfo{author}{L.~C. McInnes},
  \bibinfo{author}{K.~Rupp}, \bibinfo{author}{B.~F. Smith},
  \bibinfo{author}{S.~Zampini}, \bibinfo{author}{H.~Zhang},
  \bibinfo{author}{H.~Zhang}, \bibinfo{title}{{{PETSc Web}} page},
  \bibinfo{howpublished}{http://www.mcs.anl.gov/petsc},
  \bibinfo{year}{2017}{\natexlab{a}}.
\bibitem[{Balay et~al.(2017{\natexlab{b}})Balay, Abhyankar, Adams, Brown,
  Brune, Buschelman, Dalcin, Eijkhout, Kaushik, Knepley, May, McInnes, Gropp,
  Rupp, Sanan, Smith, Zampini, Zhang, and Zhang}]{petsc-user-ref}
\bibinfo{author}{S.~Balay}, \bibinfo{author}{S.~Abhyankar},
  \bibinfo{author}{M.~F. Adams}, \bibinfo{author}{J.~Brown},
  \bibinfo{author}{P.~Brune}, \bibinfo{author}{K.~Buschelman},
  \bibinfo{author}{L.~Dalcin}, \bibinfo{author}{V.~Eijkhout},
  \bibinfo{author}{D.~Kaushik}, \bibinfo{author}{M.~G. Knepley},
  \bibinfo{author}{D.~A. May}, \bibinfo{author}{L.~C. McInnes},
  \bibinfo{author}{W.~D. Gropp}, \bibinfo{author}{K.~Rupp},
  \bibinfo{author}{P.~Sanan}, \bibinfo{author}{B.~F. Smith},
  \bibinfo{author}{S.~Zampini}, \bibinfo{author}{H.~Zhang},
  \bibinfo{author}{H.~Zhang}, \bibinfo{title}{{{PETSc Users Manual}}},
  \bibinfo{type}{Technical Report} \bibinfo{number}{ANL-95/11 - Revision 3.8},
  {Argonne National Laboratory}, \bibinfo{year}{2017}{\natexlab{b}}.
\bibitem[{Dembo et~al.(1982)Dembo, Eisenstat, and Steihaug}]{Dembo1982}
\bibinfo{author}{R.~Dembo}, \bibinfo{author}{S.~Eisenstat},
  \bibinfo{author}{T.~Steihaug},
\newblock \bibinfo{title}{Inexact newton methods},
\newblock \bibinfo{journal}{SIAM Journal on Numerical Analysis}
  \bibinfo{volume}{19} (\bibinfo{year}{1982}) \bibinfo{pages}{400--408}.
\bibitem[{Anderson(2003)}]{Anderson2003}
\bibinfo{author}{J.~D. Anderson}, \bibinfo{title}{Modern {{Compressible Flow}}:
  {{With}} a {{Historical Perspective}}}, \bibinfo{publisher}{{McGraw-Hill New
  York}}, \bibinfo{year}{2003}.
\bibitem[{Sod(1978)}]{Sod1978}
\bibinfo{author}{G.~A. Sod},
\newblock \bibinfo{title}{A survey of several finite difference methods for
  systems of nonlinear hyperbolic conservation laws},
\newblock \bibinfo{journal}{Journal of Computational Physics}
  \bibinfo{volume}{27} (\bibinfo{year}{1978}) \bibinfo{pages}{1--31}.
\bibitem[{Woodward and Colella(1984)}]{Woodward1984}
\bibinfo{author}{P.~Woodward}, \bibinfo{author}{P.~Colella},
\newblock \bibinfo{title}{The {{Numerical Simulation}} of {{Two}}-{{Dimensional
  Fluid Flow}} with {{Strong Shocks}}},
\newblock \bibinfo{journal}{Journal of Computational Physics}
  \bibinfo{volume}{173} (\bibinfo{year}{1984}) \bibinfo{pages}{115--173}.
\bibitem[{Jasak(1996)}]{Jasak1996}
\bibinfo{author}{H.~Jasak}, \bibinfo{title}{Error {{Analysis}} and
  {{Estimation}} for the {{Finite Volume Method}} with {{Applications}} to
  {{Fluid Flow}}}, Ph.D. thesis, Imperial College London, \bibinfo{year}{1996}.
\bibitem[{Sims(1964)}]{Sims1964}
\bibinfo{author}{J.~Sims}, \bibinfo{title}{Tables for Supersonic Flow around
  Right Circular Cones at Zero Angle of Attack}, \bibinfo{type}{Technical
  Report} \bibinfo{number}{NASA-SP-3004}, {NASA Marshall Space Flight Center},
  \bibinfo{address}{Huntsville, AL, USA}, \bibinfo{year}{1964}.
\bibitem[{Kutler and Lomax(1971)}]{Kutler1971}
\bibinfo{author}{P.~Kutler}, \bibinfo{author}{H.~Lomax},
\newblock \bibinfo{title}{A systematic development of the supersonic flow
  fields over and behind wings and wing-body configurations using a
  shock-capturing finite-difference approach},
\newblock \bibinfo{publisher}{{AIAA 9th Aerospace Science Meeting, AIAA Paper
  No. 71-99}}, \bibinfo{year}{1971}.

\end{thebibliography}
\end{document}